\documentclass{article}

\usepackage{amsmath}
\usepackage{graphicx}
\usepackage{subcaption}
\usepackage{float}
\usepackage{wrapfig}
\usepackage{enumerate}
\usepackage{booktabs}
\usepackage{tabularx}
 \PassOptionsToPackage{numbers, compress}{natbib}
     \usepackage[preprint]{neurips_2025}

\usepackage[utf8]{inputenc} 
\usepackage[T1]{fontenc}    
\usepackage{hyperref}       
\usepackage{url}            
\usepackage{booktabs}       
\usepackage{amsfonts}       
\usepackage{nicefrac}       
\usepackage{microtype}      
\usepackage{xcolor}         

\newcommand{\appropto}{\mathrel{\vcenter{
  \offinterlineskip\halign{\hfil$##$\cr
    \propto\cr\noalign{\kern2pt}\sim\cr\noalign{\kern-2pt}}}}}
\title{Contrastive Normalizing Flows for Uncertainty-Aware Parameter Estimation}

\author{%
  Ibrahim Elsharkawy \\
  Department of Physics\\
  University of Illinois Urbana-Champaign\\
  Urbana, IL, USA \\
  \texttt{ie4@illinois.edu} \\
     \And
  Yonatan Kahn \\
  Department of Physics, University of Toronto\\
  and Vector Institute\\
  Toronto, ON, Canada \\
  \texttt{yf.kahn@utoronto.ca} \\
}

\begin{document}

\maketitle

\begin{abstract}
Estimating physical parameters from data is a crucial application of machine learning (ML) in the physical sciences. However, systematic uncertainties, such as detector miscalibration, induce data distribution distortions that can erode statistical precision. In both high-energy physics (HEP) and broader ML contexts, achieving uncertainty-aware parameter estimation under these domain shifts remains an open problem. In this work, we address this challenge of uncertainty-aware parameter estimation for a broad set of tasks critical for HEP. We introduce a novel approach based on Contrastive Normalizing Flows (CNFs), which achieves top performance on the HiggsML Uncertainty Challenge dataset. Building on the insight that a binary classifier can approximate the model parameter likelihood ratio, we address the practical limitations of expressivity and the high cost of simulating high-dimensional parameter grids by embedding data and parameters in a learned CNF mapping. This mapping yields a tunable contrastive distribution that enables robust classification under shifted data distributions. Through a combination of theoretical analysis and empirical evaluations, we demonstrate that CNFs, when coupled with a classifier and established frequentist techniques, provide principled parameter estimation and uncertainty quantification through classification that is robust to data distribution distortions.
\end{abstract}

\section{Introduction}
A critical application of machine learning (ML) in the physical sciences is parameter estimation. Given a model of a physical system (for example, a high-energy collider experiment) that depends on certain parameters (such as the masses and interaction strengths of the colliding particles), it is generally straightforward to perform forward modeling and make predictions for experiments. However, the inverse problem --- reliably estimating model parameters from observed data --- is generically much more difficult. A key complication of this inverse problem is the combined effect of systematic error, such as calibration inaccuracies or variations in some measurement apparatus's response, which induce deformations in the observed data. These systematic uncertainties can be modeled with nuisance parameters in the data distribution and must either be estimated or marginalized over to extract the parameters of interest.

In a broad set of problems in high-energy physics (HEP), the parameter of interest is the fraction of events drawn from a signal distribution in a mixture of signal and background. A precise signal‐fraction estimate enables the measurement of physical parameters (like coupling strengths or particle masses) and the reliable assessment of a particle discovery significance, given proper uncertainty quantification. To address this need, we present a novel method to extract the signal fraction of a dataset with principled uncertainty estimates from data deformed by systematic error. The key features of this method are (a) the use of \emph{contrastive normalizing flows} (CNFs) to both model and distinguish the signal and background distributions, and (b) a deep neural network (DNN) classifier which takes as input the learned CNF probabilities evaluated on data perturbed by a variety of nuisance parameters. We demonstrate that our approach is robust to the effects of nuisance parameters perturbing the signal/background boundary, and furthermore, that it can provide reliable confidence intervals on the signal strength parameter even in the presence of unknown systematic error. The potential of this method is demonstrated by its performance on the recent HiggsML Uncertainty Challenge dataset, the first benchmark for uncertainty-aware parameter estimation methodologies in HEP \cite{bhimji2024fairuniversehiggsmluncertainty}. Our specific contributions can be summarized as follows: 
\begin{itemize}
    \item We apply CNFs to binary classification under domain shift task and develop a novel loss, architecture, and training procedure to ensure stable and accurate learning (Sec.~\ref{sec:CNF}).
    \item  Using an analysis of the contrastive loss, we show that on a toy problem, classifiers trained on CNF features provide robust classification accuracy under domain shift (Sec.~\ref{sec:toy}).
    \item We develop a novel parameter estimation method with CNFs and a DNN classifier to estimate the signal fraction with accurate $1\sigma$ confidence intervals under nuisance-parameter induced systematic shifts (Sec.~\ref{sec:challenge}).
    \item We demonstrate the efficacy of the CNF-based parameter estimation method on the Higgs ML Uncertainty Challenge dataset and show that it outperforms alternative approaches.
\end{itemize}

\section{Related Works}
\label{sec:related_works}
Systematic uncertainties in HEP describe distribution mismatches between simulation (training) and real data, and thus are directly analogous to domain shifts in ML. \cite{louppe2020adversarialvariationaloptimizationnondifferentiable,ovadia2019trustmodelsuncertaintyevaluating,pmlr-v139-koh21a}. Techniques such as invariant risk minimization and distributionally robust optimization \cite{arjovsky2020invariantriskminimization,sagawa2020distributionallyrobustneuralnetworks} or adversarial domain adaptation and uncertainty-aware inference \cite{malinin2018predictiveuncertaintyestimationprior,lipton2017mythosmodelinterpretability} have direct counterparts in HEP —e.g.\ adversarial pivots and end-to-end optimization—to mitigate nuisance effects \cite{Baldi_2016,louppe2017learningpivotadversarialnetworks,Ghosh_2021,dorigo2022endtoendoptimizationparticlephysics}. Although numerous techniques exist across both fields, each entails different assumptions and tradeoffs. Methods that enforce invariance to nuisance parameters often sacrifice raw discriminative power, whereas methods that condition explicitly on nuisance parameters must trust the precise form of the parameterization~\cite{Baldi_2016, louppe2017learningpivotadversarialnetworks, Ghosh_2021}. Empirically, no single approach has reliably emerged as optimal for systematically perturbed data. This is a primary reason for the development of the Higgs ML challenge, which could serve as a benchmark for such methods in the HEP community and possibly the broader ML community \cite{bhimji2024fairuniversehiggsmluncertainty}. 

Normalizing flows (NFs) ~\cite{dinh2017densityestimationusingreal,kingma2018glowgenerativeflowinvertible, papamakarios2021normalizingflowsprobabilisticmodeling} offer a powerful generative framework capable of density estimation and invertible transformations. However, it has been shown that flows are unreliable for out-of-distribution detection if used naively~\cite{nalisnick2019detectingoutofdistributioninputsdeep}. In response, a line of work has proposed using multiple learned distributions to improve normalizing flow performance for anomaly detection ~\cite{ren2019likelihoodratiosoutofdistributiondetection, schmier2023positivedifferencedistributionimage}. Notably, CNFs~\cite{schmier2023positivedifferencedistributionimage} have been shown to emphasize density differences between in-and out-of-distribution data, mitigating a known pitfall of likelihood-based anomaly detection. In parallel, generative-discriminative hybrid models~\cite{9879060} illustrate the promise of invertible densities with classification objectives to handle distribution shifts. On the HEP side, this manifests as simulation-to-data reweighting (i.e.\ OmniFold~\cite{Andreassen_2020}), adversarial training to remove nuisance parameter dependence~\cite{louppe2017learningpivotadversarialnetworks}, and network parameterization to incorporate uncertain physics~\cite{Baldi_2016, Ghosh_2021}. 

Our approach aims to merge these two lines of work by introducing a method using CNFs designed for robust parameter estimation under systemic shifts. By leveraging a contrastive training objective, we learn to disentangle relevant features from systematic variations, producing well-calibrated parameter estimates in the presence of domain shifts.

\section{Problem Setup and Motivation}
\label{sec:setup}

\paragraph{Binary classifiers for parameter estimation} For any general estimation problem, a \textit{binary classifier} has been shown to be capable of parameter estimation in the limit of infinite data and infinite expressivity \cite{cranmer2016approximatinglikelihoodratioscalibrated}. Suppose  $\mathbf{x} \sim P(\mathbf{x}|\{\Theta_i,\nu_i\})$ where $\{\Theta_i\}$ are a set of model parameters to be estimated and $\{\nu_i\}$ are a set of nuisance parameters. One can feed the classifier this event $\mathbf{x}$, the parameters it was drawn from $\{\Theta_i,\nu_i\}$ and some other set of incorrect parameters $\{\Theta'_i,\nu'_i\}$, with the binary classification task of identifying which set of parameters $\mathbf{x}$ was drawn from. Given many different observations $\{\mathbf{x}_i \}$ drawn from many different choices of parameters, the classifier will learn to be monotonic in the likelihood ratio of any arbitrary combination of parameters:  
\begin{equation}
r(\mathbf{x},\{\Theta_i,\nu_i\},\{\Theta'_i,\nu'_i\})\appropto\frac{P(\mathbf{x}|\{ \Theta_i,\nu_i\})}{P(\mathbf{x}|\{ \Theta'_i,\nu'_i\})}.
\label{eq:cranmerliklihood}
\end{equation}
Maximum likelihood estimation can thus be used to find the most probable set of parameters along with uncertainty estimates \cite{cranmer2016approximatinglikelihoodratioscalibrated}. 

Despite this theoretical guarantee, in many practical cases, classifiers empirically fall short for parameter estimation. If the number of model parameters $k_\Theta$ or nuisance parameters $k_\nu$ is large, the curse of dimensionality prevents sufficient sampling of parameter space, as generating a training set for each choice of $\{\Theta_i,\nu_i\}$ may be prohibitively expensive. In addition, the effect of the model and nuisance parameters on individual data points is often minor.

 \paragraph{Problem setup} We focus on a situation where data $\mathbf{x} \in \mathbb{R}^m$ is drawn from a mixture of two distributions, a \emph{signal} distribution $p_s(\mathbf{x} |\{\nu_i \})$ and a \emph{background} distribution $p_b(\mathbf{x} |\{\nu_i \})$, where both distributions may depend on a set of unknown nuisance parameters $\{\nu_i \}$ parameterizing some systematic error. The task is to estimate some model parameter $\Theta$ proportional to the signal fraction $f_s \equiv \frac{N_s}{N_s + N_b}$ where $N_s$ and $N_b$ are the number of signal and background events in some test set. In many cases of relevance, $f_s$ is much smaller than 1. In the HiggsML Uncertainty Challenge, the motivating example for our method, $f_s\sim10^{-4}-10^{-3}$, and the rarity of the signal is such that a $20\%$ increase in $f_s$ may lead to an increase of fewer than 20 signal events in a dataset of $10^6$ events~\cite{bhimji2024fairuniversehiggsmluncertainty}. Thus, a classifier must learn to detect the subtle effect of a 20\% increase in $f_s$ by training on individual, unlabeled events drawn from many large datasets with varying $f_s$ and $\nu$. 
 
\paragraph{Signal fraction parameter estimation} In this case where $\Theta\propto f_s$, we can simplify the classifier task by replacing model parameters $\Theta_i,\Theta'_i$ in Eq.~(\ref{eq:cranmerliklihood}) with signal and background labels.
That is, we train a classifier to distinguish signal events drawn from $p_s(\mathbf{x} | \{\nu_i\})$ from background events drawn from $p_b(\mathbf{x} | \{\nu'_i\})$. In the limit of infinite data and expressivity, this classifier will approximate 
\begin{equation}
r(\mathbf{x}, \{\nu_i\}, \{\nu'_i\})\appropto\frac{p_s(\mathbf{x}|\{\nu_i\})}{p_{b}(\mathbf{x}|\{\nu'_i\})} \label{eq:nophilikli}
\end{equation}
and given proper calibration, this likelihood ratio can be used to estimate $\Theta\propto f_s$. To remedy the issue of high dimensionality $k_\nu \gg 1$, the potential degeneracies between nuisance parameters $\nu_i$ and parameters distinguishing $p_s$ and $p_b$, and the need for high precision in estimation, we propose the introduction of signal and background \emph{discrimination functions} $\Phi_{s,b}[\mathbf{x};\{\nu_i\}]$, which implicitly depend on the nuisance parameters $\nu_i$ through their effect on $p_{s,b}(\mathbf{x} | \{\nu_i\})$. Using $\Phi_{s,b}$ as inputs to the classifier instead of the raw nuisance parameters, the classifier approximates
\begin{equation}
r(\mathbf{x}, \{\nu_i\}, \{\nu'_i\})\appropto\frac{p_s(\mathbf{x}|\Phi_s[\mathbf{x};\{\nu_i\}])}{p_{b}(\mathbf{x}|\Phi_b[\mathbf{x};\{\nu'_i\}])}.
\label{eq:philikli}
\end{equation}
If $\Phi_{s}[\mathbf{x};\{\nu_i\}]$ takes very different values when $\mathbf{x} \sim p_s$ compared to $\mathbf{x} \sim p_b$ (and likewise for $\Phi_b$), \textit{and} this discrimination ability is relatively unaffected by varying the nuisance parameters $\{\nu_i\}$, the task of approximating the likelihood ratio in Eq.~(\ref{eq:philikli}) will be easier than learning either the full likelihood ratio in Eq.~(\ref{eq:cranmerliklihood}) or the specialized likelihood ratio in Eq.~(\ref{eq:nophilikli}), allowing data-efficient training.

\paragraph{Motivation for CNFs} In this work, we choose the discrimination functions $\Phi_{s,b}$ to be monotonic functions of the learned probabilities of CNFs trained on \emph{unperturbed} data, and take the classifier $r$ to be a DNN trained on \emph{perturbed} data drawn from distributions where the nuisance parameters take a variety of values. As we will demonstrate, the robustness of the learned CNF distributions to the nuisance parameters allows successful classification with just 1000 choices of nuisance parameters from a fairly high-dimensional set $k_\nu = 6$, efficiently sampling the nuisance parameter space with just $1000^{1/6} \approx 3$ choices of each nuisance parameter value.

\section{Contrastive Normalizing Flows}
\label{sec:CNF}
We begin by describing the setup of traditional NFs, and then introduce our modification to obtain CNFs. The standard loss used to train NFs maximizes the log-likelihood of a set of data $\mathcal{D}$,
\begin{equation}
    \mathcal{L} =  \frac{1}{|\mathcal{D}|}\sum_{\mathbf{x}\in\mathcal{D}} -\log(p_\theta(\mathbf{x}))  \equiv \frac{1}{|\mathcal{D}|} \sum_{\mathbf{x}\in\mathcal{D}} -\Big [\log p_z(f_\theta(\mathbf{x})) + \log \left| \det J_{f_\theta}(x) \right| \Big],
\end{equation}
where $p_z(\mathbf{x})$ is the base distribution of the NF (usually taken to be a multivariate Gaussian $\mathcal{N}(0,I_{m \times m})$), $f_\theta(\mathbf{x})$ is the transformation with learnable parameters $\theta$ which maps $p_z$ to the learned distribution $p_\theta$, and the Jacobian of the transformation is $J_{f_\theta}$. 

In our case, $\mathcal{D}$ is a mixture of signal and background events such that $\mathcal{D} = \{(\mathbf{x}_s,\mathbf{x}_b)~|~ \mathbf{x}_s \sim p_s(\mathbf{x}),\mathbf{x}_b \sim p_b(\mathbf{x})\}$. To learn a discrimination function which can distinguish between the two distributions $p_s$ and $p_b$, we propose the introduction of a contrastive term in the loss:
\begin{equation}
\begin{split}
\label{eq:cnfloss}
        &\mathcal{L}_s = \frac{1}{|\mathcal{D}|} \sum_{\mathcal{D}} \Big\{-\log p_\theta^{(s)}(\mathbf{x}_s) +c\,\log p_\theta^{(s)}(\mathbf{x}_b) \Big\} \\
       &\hspace{-2mm}\equiv \frac{1}{|\mathcal{D}|} \sum_{\mathcal{D}}\bigg\{ - \big[ \log p_z(f_\theta(\mathbf{x}_s)) 
+ \log \big| \det J(\mathbf{x}_s) \big| \big]  + c \big[ \log p_z(f_\theta(\mathbf{x}_b)) 
+ \log \big| \det J(\mathbf{x}_b) \big| \big] 
\bigg\}
\end{split}
\end{equation}
where $c\in\mathbb{R}^+$ is the hyperparameter governing the strength of the contrastive loss, and $p_\theta^{(s)}(\mathbf{x})$ is the learned distribution. The training objective which defines the CNF for $\mathcal{L}_s$ is to simultaneously maximize the log-likelihood of $\mathcal{S}$ while minimizing the log-likelihood of $\mathcal{B}$ on training data with signal/background labels. Note that $\mathcal{L}_s$ breaks the symmetry between signal and background; we would obtain a different learned distribution $p_\theta^{(b)}$ from the \emph{same} dataset $\mathcal{D}$ by reversing the roles of $\mathbf{x}_s$ and $\mathbf{x}_b$ in the loss function, which we would denote $\mathcal{L}_b$ when $c$ penalizes $\mathbf{x}_s$ rather than $\mathbf{x}_b$. In the notation of Sec.~\ref{sec:setup}, we will take the discrimination functions to be $\Phi_s(\mathbf{x}) = \frac{p_\theta^{(s)}(\mathbf{x})}{1+p_\theta^{(s)}(\mathbf{x})}$ and $\Phi_b(\mathbf{x}) = \frac{p_\theta^{(b)}(\mathbf{x})}{1+p_\theta^{(b)}(\mathbf{x})}$, which are monotonic in $p^{(s,b)}$ but bounded between 0 and 1.

In practice, training a CNF can be challenging due to mode collapse and sharp boundaries in the contrastive distribution. We develop a specialized training procedure and architecture, which we find empirically \textit{necessary} to achieve accurate and stable learning, and are described in detail in App. \ref{app:ArchCNF}.

\subsection{Analytic minimization of the contrastive loss}
\label{sec:analysis_loss}
In the limit of infinite data and an equal mixture of signal and background in $\mathcal{D}$, our loss can be approximated via integration over the signal and background distributions:
\begin{align}
    \mathcal{L}_s & = \frac{1}{2} \Big [-\,\mathbb{E}[\log p_\theta^{(s)}(\mathbf{x})]+c\,\mathbb{E}[\log p_\theta^{(s)}(\mathbf{x})] \Big ] \\
    & = \frac{1}{2} \left [-\int p_s(\mathbf{x})\,\log p_\theta^{(s)}(\mathbf{x})\, d\mathbf{x} +c\int p_b(\mathbf{x})\,\log p^{(s)}_\theta(\mathbf{x})\,d\mathbf{x} \right ].
\end{align}
 Minimizing the loss becomes a functional optimization problem for the learned distribution $p^{(s)}_\theta(\mathbf{x})$:
\begin{equation}
\min_{p^{(s)}_\theta(\mathbf{x}) \ge 0}\Bigl(-\int p_s(\mathbf{x})\,\log p_\theta^{(s)}(\mathbf{x})\, d\mathbf{x}_s +c\int p_b(\mathbf{x})\,\log p^{(s)}_\theta(\mathbf{x})\,d\mathbf{x} \Bigr) \quad \text{subject to} \quad \int p_\theta(\mathbf{x})\,d\mathbf{x} = 1.
\end{equation}
Imposing the constraint with a Lagrange multiplier $\lambda$ entails the minimization of the functional
\begin{equation}
\mathbf{L}\bigl[p^{(s)}_\theta(\mathbf{x}),\lambda\bigr]=-\lambda +\, \int d\mathbf{x} \left\{ (-p_s(\mathbf{x}) + c p_y(\mathbf{x}))\,\log p^{(s)}_\theta(\mathbf{x})+ \lambda \, p^{(s)}_\theta(\mathbf{x})\right\}.
\end{equation}
Taking the functional derivative with respect to $p^{(s)}_\theta$, we find the learned distribution that minimizes the loss,
\begin{equation}
\label{eq:opt}
    p^{(s)}_{\theta^*}(\mathbf{x}) = \frac{1}{\lambda_c}\text{max}\Big\{[ p_s(\mathbf{x})-c\, p_b(\mathbf{x})],0\Big\},
\end{equation}
where $\lambda_c$ is now interpreted as a ($c$-dependent) normalization factor for $p^{(s)}_{\theta^*}(\mathbf{x})$. By construction, $p^{(s)}_{\theta^*}(\mathbf{x})$ vanishes in any portion of data-space where the likelihood ratio is less than $c$: $\frac{p_s(\mathbf{x})}{p_b(\mathbf{x})} \leq c$. Eq.~(\ref{eq:opt}) represents a contrastive distribution, with the degree of contrast between signal and background controlled by $c$. A value of $c=1$ ensures that the boundary where $p^{(s)}_{\theta^*}(\mathbf{x})$ vanishes, $c\, p_b(\mathbf{x})= p_s(\mathbf{x})$, corresponds to the unit contour of the log-likelihood ratio.

\subsection{CNFs for robust classification}
We now explore why the CNF's learned probabilities are suitable for constructing discrimination functions $\Phi_{s,b}[\mathbf{x};\{\nu_i\}]$ in Eq.~(\ref{eq:philikli}). Compared to a classifier, which learns a decision boundary between two classes, the CNF models a probability distribution $p^{(s,b)}_{\theta^*}(\mathbf{x})$ which contains information about the distribution of each class. The boundary where $p^{(s,b)}_{\theta^*}(\mathbf{x})$ vanishes corresponds to a contour of the likelihood ratio (the Bayes-optimal discrimination boundary for a binary classifier), but the nonvanishing parts of the distribution also capture the underlying structure of each class. Tuning the hyperparameter $c$ redistributes the learned probability mass to emphasize certain class features differently. This is distinct from simply reweighting classes or choosing different score cuts in a binary cross-entropy (BCE) classifier, where only the decision boundary is shifted but the learned feature representation is essentially unchanged. Due to this holistic view of the data distribution, we expect CNFs to be more robust to systematic uncertainties. If correlations among the input features shift due to nuisance parameters, a model that has learned the entire distribution is less likely to fail catastrophically than one that focuses primarily on class boundaries. We validate this intuition on a toy example in the following Sec.~\ref{sec:toy}, and demonstrate the efficacy of our approach on the full HiggsML Uncertainty Challenge in Sec.~\ref{sec:challenge}. Another major difference between CNFs and a classifier is the CNFs \textit{generative capability}. Although not used for the method, CNFs provide a method to generate samples from one distribution most unlike another (controllably with $c$). We provide visual examples of CNFs trained on MNIST for classification and generative cases in Appendix \ref{app:mnist}.

\begin{figure}[t]
  \centering
    \includegraphics[width=0.33\linewidth]{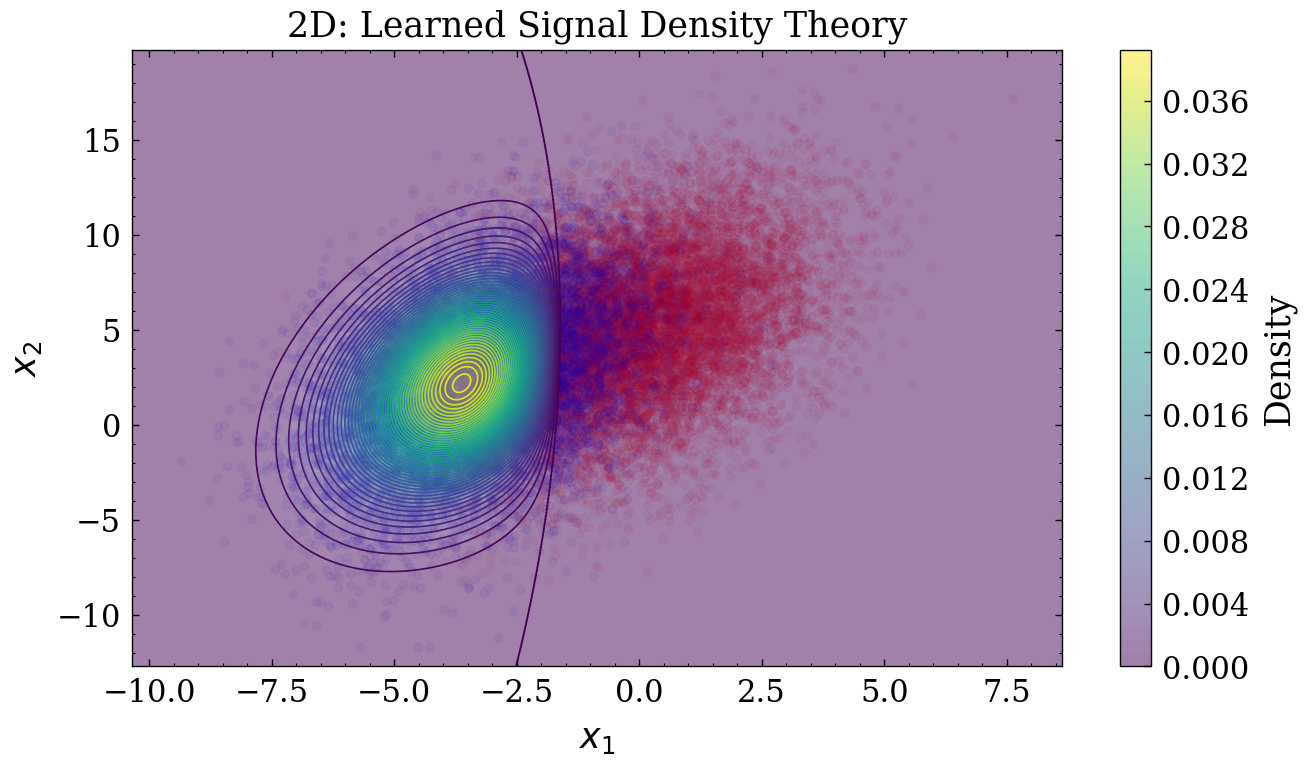}
    \includegraphics[width=0.33\linewidth]{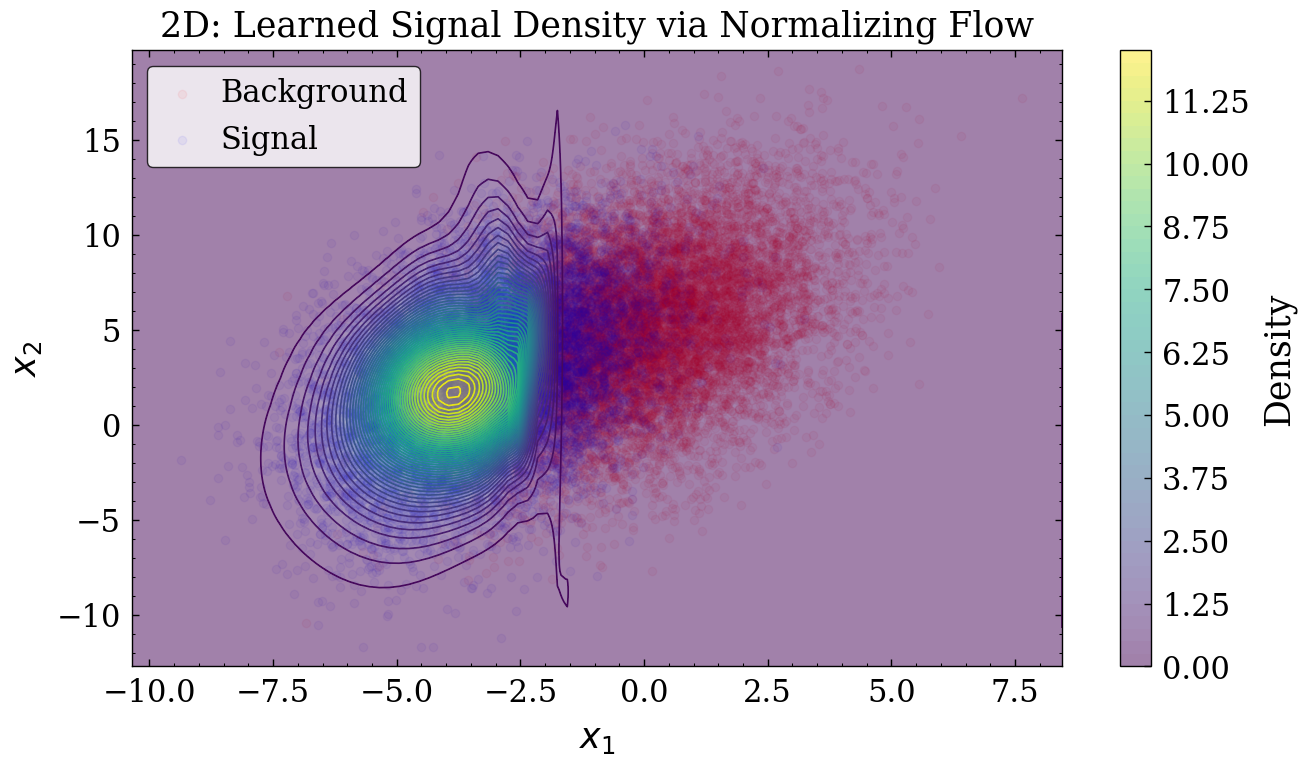}
    \includegraphics[width=0.32\linewidth]{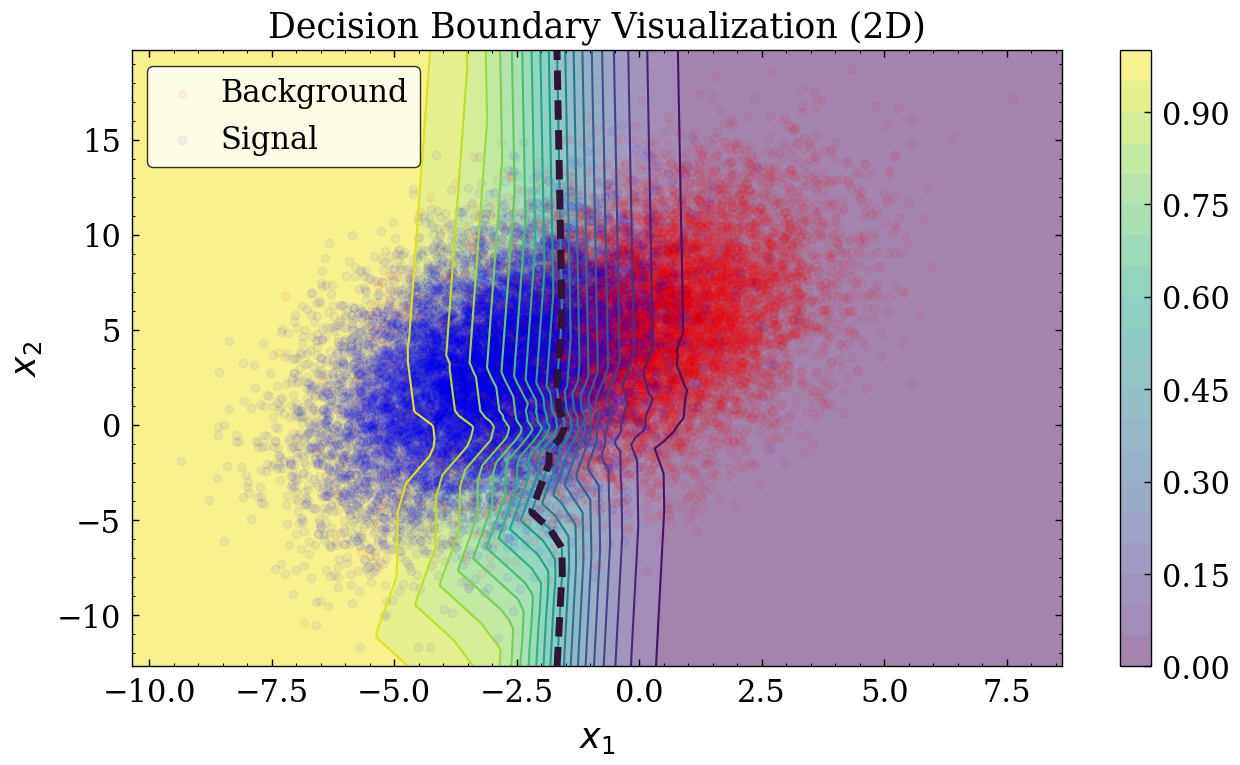}
  \caption{Randomly sampled signal (blue) and background (red) points with contours overlaid. CNF loss-minimizing distribution contour $p^{(s)}_{\theta^*}(\mathbf{x})$ for $c =1$ (left), learned CNF distribution contour for $c=1$ (center), and DNN classifier score contour (right) for two 2-dimensional Gaussian distributions. The DNN decision boundary $r = 0.5$ (dashed black, right panel) closely matches the zero contour of the CNF distributions.}
  \label{fig:toy_prob}
  \vspace{-0.3cm}
\end{figure}

\section{Toy Example: Two Gaussian Distributions}
\label{sec:toy}

To illustrate the efficacy of CNFs, we train a CNF on a toy problem of data drawn from two Gaussian distributions, one we call signal and one we call background. For ease of visualization, we start with 2-dimensional Gaussians and later generalize to higher dimensions.

Fig.~\ref{fig:toy_prob} shows 30,000 samples from $p_s = \mathcal{N}(\boldsymbol{\mu}_s, \boldsymbol{\Sigma}_s)$ (blue) and $p_b = \mathcal{N}(\boldsymbol{\mu}_b, \boldsymbol{\Sigma}_b)$ (red) where
\begin{equation}
\boldsymbol{\mu}_s = \begin{bmatrix}-3.25 \\2.50\end{bmatrix},~\boldsymbol{\Sigma}_s =\begin{bmatrix}2.58 & 2.19 \\2.19 & 13.06\end{bmatrix}\quad\boldsymbol{\mu}_b = \begin{bmatrix}-0.09 \\5.00\end{bmatrix},~\boldsymbol{\Sigma}_b =\begin{bmatrix}3.30 & 2.63 \\ 2.63 & 14.23\end{bmatrix}.
\label{eq:musigma2}
\end{equation}
The left panel shows a density plot of the loss-minimizing CNF distribution $p_{\theta^*}^{(s)}$ from Eq.~(\ref{eq:opt}) for $c=1$. The sharp cutoff $p_{\theta^*}^{(s)} = 0$ is visible for points sufficiently far from $\mu_s$ in the direction of $\mu_b$. The center panel shows the same samples with a density plot of $p_{\theta}^{(s)}$ as learned by a trained CNF with $c = 1$ (with the architecture described in App. \ref{app:ArchCNF}), which matches our analytic expectations well. Finally, we train a 3-layer, width-32 DNN classifier with ReLU activation and cross-entropy loss on an equal mixture of samples, and overlay a density plot of the classifier score $r$ in the right panel. We see the DNN decision boundary $r = 0.5$ matches the boundary $p_\theta^{(s)} = 0$ learned by the $c=1$ CNF, as expected. Appendix \ref{app:vary_c} shows the results of training with different values of $c$.

\begin{figure}[t]
  \centering
    \includegraphics[width=0.30\linewidth]{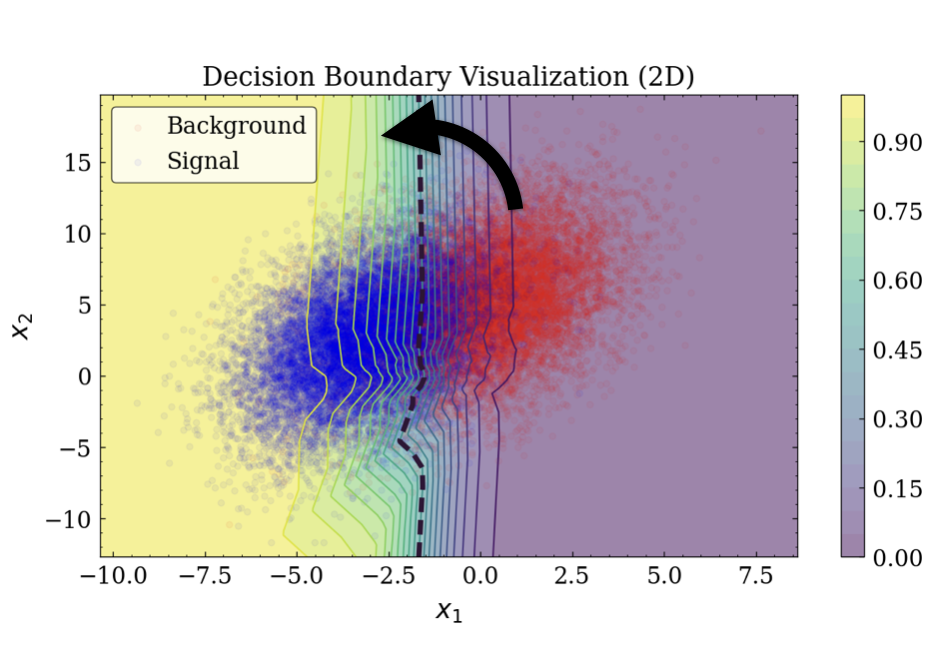}
    \includegraphics[width=0.31\linewidth]{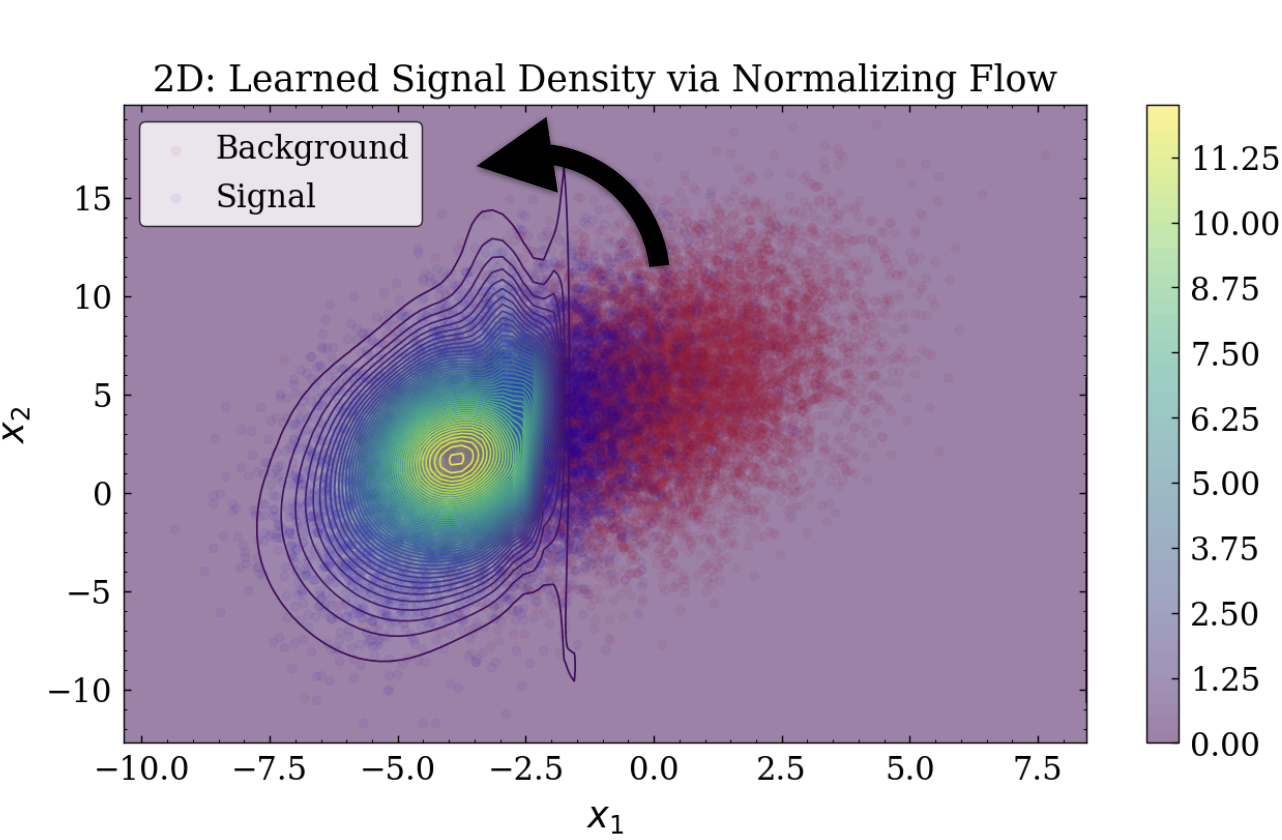}
  \caption{A rotation of the data (black arrows) can move points across the DNN decision boundary (dashed black, left), but remain in the $p_{\theta}^{(s)} = 0$ region of the CNF distribution (right).}
  \label{fig:cnfdnnrot}
\end{figure}

With this example in mind, suppose we distort the data with a rotation in the $x_1-x_2$ plane, as illustrated in Fig.~\ref{fig:cnfdnnrot}. It is clear that the DNN decision boundary (left), which extends well beyond the region of high signal or background density, is more prone to mis-classification under this distortion; for the DNN classifier, a background point is rotated into the signal region, while for the CNF (right), the background is still in a background region with $p_\theta^{(s)} \approx 0$ after rotation. To validate this intuition, we generalize to a 10-dimensional example. We construct \[
\boldsymbol\mu_s^{\text{ext}} =
\begin{bmatrix}
\boldsymbol{\mu}_s \\
 \mathbf{0}_8
\end{bmatrix},
\quad
\boldsymbol\mu_b^{\text{ext}} =
\begin{bmatrix}
\boldsymbol{\mu_b} \\
 \mathbf{0}_8
\end{bmatrix},\quad
\boldsymbol\Sigma_s^{\text{ext}} =
\begin{bmatrix}
\boldsymbol{\Sigma}_s & 0 \\
0 & BB^\top
\end{bmatrix},
\quad
\boldsymbol{\Sigma}_b^{\text{ext}} =
\begin{bmatrix}
\mathbf{\Sigma}_b & 0 \\
0 & BB^\top
\end{bmatrix}
\]
with $\boldsymbol{\mu}_{s,b}$ and $\boldsymbol{\Sigma}_{s,b}$ as in Eq.~(\ref{eq:musigma2}), but  $\boldsymbol{\mu}_{s,b}^{\text{ext}} \in \mathbb{R}^{10}$ and $B$ a random $8 \times 8$ matrix with entries drawn from $\mathcal{N}(0,1)$. We then choose a random $10 \times 10$ matrix $A$ with the same Gaussian distribution of entries and perform a QR decomposition $A = QR,~ Q \in \mathrm{O}(10)$ so that our final 10-dimensional signal and background distributions have mean and variance given by
\begin{equation}
\boldsymbol\mu_s^{\text{final}} = Q \boldsymbol\mu_s^{\text{ext}}, \quad
\boldsymbol\mu_b^{\text{final}} = Q \boldsymbol\mu_b^{\text{ext}},\quad
\boldsymbol\Sigma_s^{\text{final}} = Q \boldsymbol\Sigma_s^{\text{ext}} Q^\top, \quad
\boldsymbol\Sigma_b^{\text{final}} = Q \boldsymbol\Sigma_b^{\text{ext}} Q^\top.
\label{eq:10Ddef}
\end{equation}
We study two types of nuisance parameters: 
\begin{enumerate}
    \item A rotation of the data in the 2D subspace (applying $R(\phi) \in {\rm SO}(2)$ to $\boldsymbol{\mu_s},\boldsymbol{\mu_b},\boldsymbol{\Sigma}_s~,\boldsymbol{\Sigma}_b$) 
    \item A continuous deformation of the embedding subspace, by generating another $10 \times 10$ random Gaussian matrix $\widetilde{A}$ and taking the QR decomposition of $A + \phi \widetilde{A}$ to obtain $Q$ in Eq.~(\ref{eq:10Ddef})
\end{enumerate}
 where the size of the deformation is parameterized by nuisance parameter $\phi$. For both scenarios, we train four networks, and for scenario (1.) we compute three corresponding predictions as described in Tab. \ref{tab:networks-training-theory}.

\begin{table}[t]
  \caption{Four approaches to the 10D toy problem; experimental results are shown in Fig.~\ref{fig:Accvstheta}.}
      \label{tab:networks-training-theory}
  \vskip 0.15in
  \begin{center}
    \small
    \setlength{\tabcolsep}{4pt}
    \begin{tabularx}{\linewidth}{@{} l 
        >{\raggedright\arraybackslash}X 
        >{\raggedright\arraybackslash}X @{}}
      \toprule
      \textbf{Network} & \textbf{Training Setup} & \textbf{Predicted Performance} \\
      \midrule
      (i) DNN classifier 
        & Unperturbed data ($\phi=0$) 
        & Bayes‐optimal classifier \\[6pt]

      (ii) CNF ($c=1$) 
        & Unperturbed data to learn $p^{(s)}_\theta(\mathbf{x})$ 
        & Loss‐minimizing $p^{(s)}_{\theta^*}(\mathbf{x})$ (Eq.~\ref{eq:opt}) \\[6pt]

      (iii) DNN classifier 
        & 10,000 perturbed samples (100 per $\phi$ across grid) 
        & Full likelihood ratio with ${\nu_i}$ (Eq.~\ref{eq:nophilikli})\\[6pt]

      (iv) DNN with $\Phi_{s,b}(\mathbf{x})$ 
        & Perturbed data \emph{and} $\Phi_{s,b}$ from (ii) 
        & Full likelihood ratio with ${\nu_i}$ (Eq.~\ref{eq:nophilikli}) \\
      \bottomrule
    \end{tabularx}
  \end{center}
\end{table}
\begin{figure}[t]
  \centering
    \includegraphics[width=0.32\linewidth]{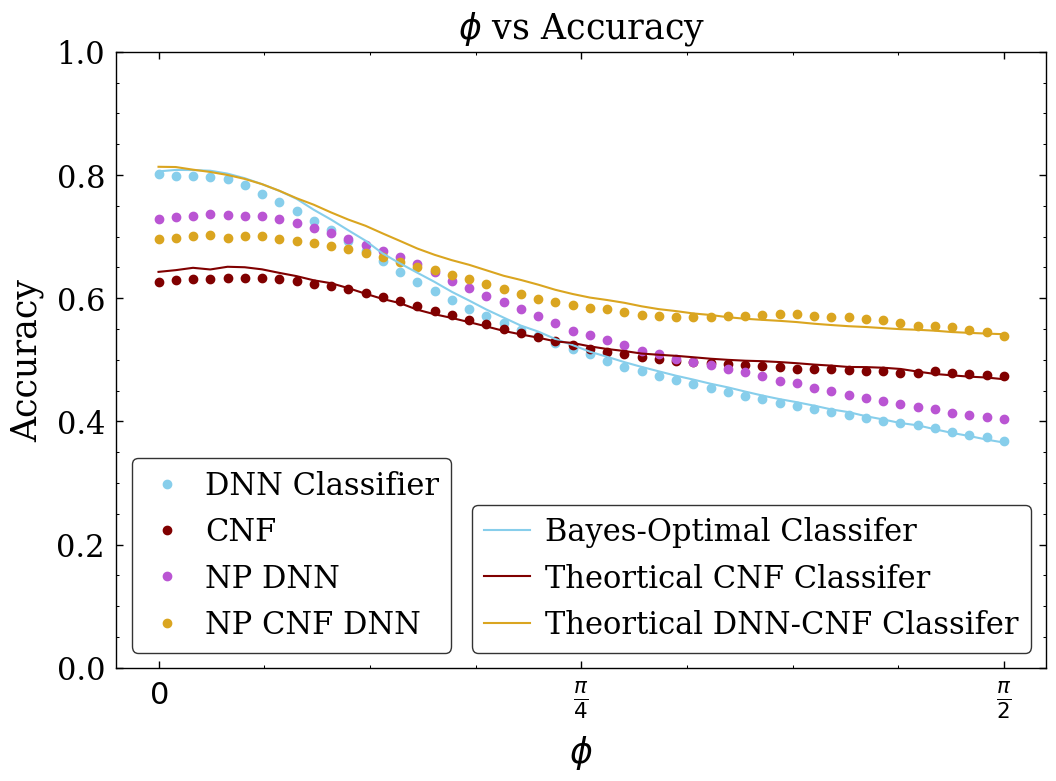}
    \includegraphics[width=0.32\linewidth]{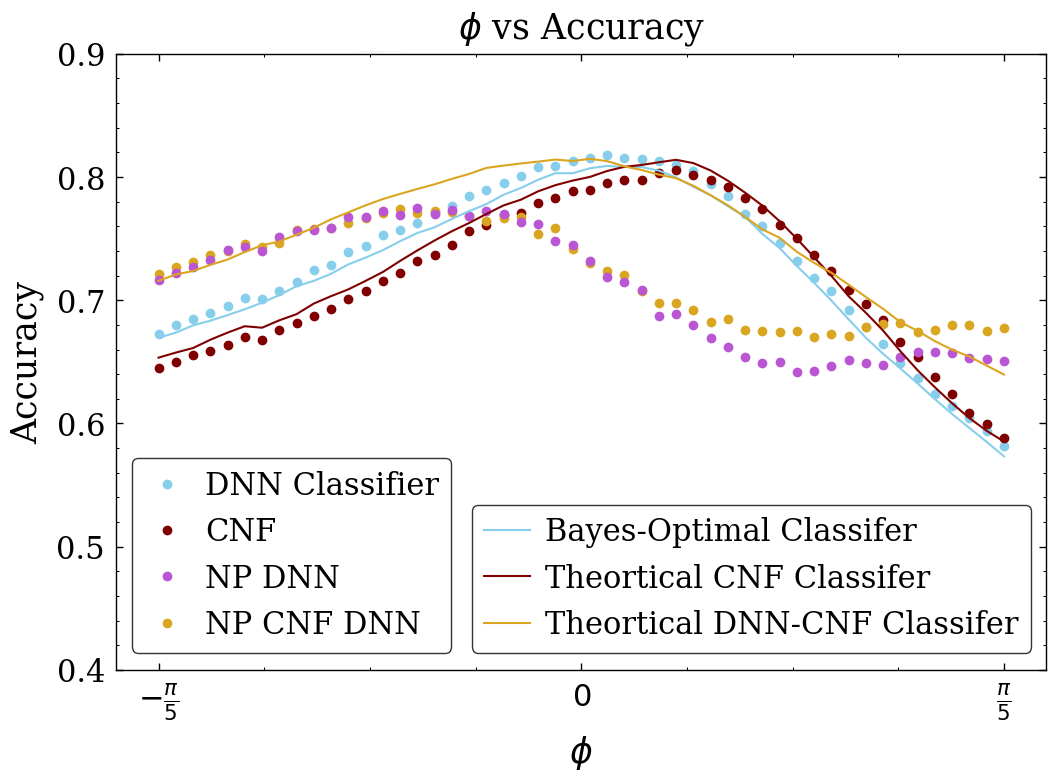}
    \includegraphics[width=0.32\linewidth]{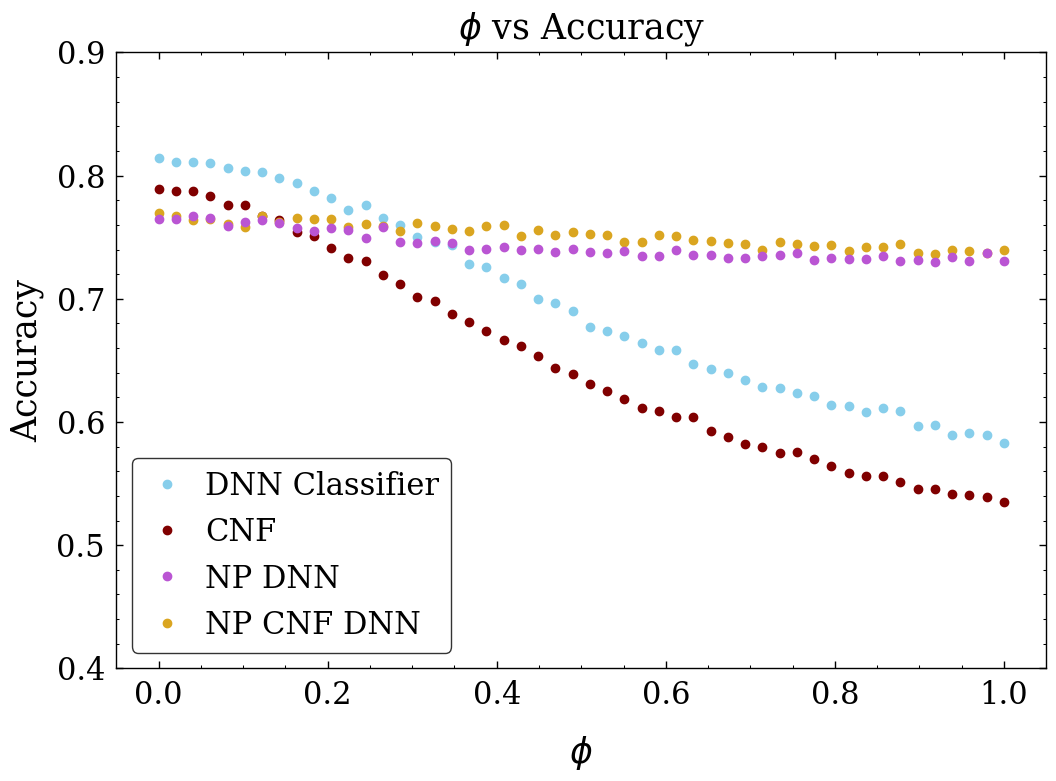}
    \label{fig:Accvstheta_subspace}
  \caption{Accuracy as a function of nuisance parameter $\phi$ for three different types of nuisance parameter deformations: a large 2D subspace rotation (left), a small 2D subspace rotation (center), and a subspace deformation (right). See text for details.}
  \label{fig:Accvstheta}
  \vspace{-0.3cm}
\end{figure}

From our arguments in Sec.~\ref{sec:setup}, we expect case (iv) to be the most robust to deformations of the data as it should approximate the full likelihood ratio including nuisance parameters. To evaluate performance, in Fig.~\ref{fig:Accvstheta} we plot the four networks' accuracies (fraction of events correctly classified) as a function of the nuisance parameter $\phi$ which characterizes the test set, setting the decision boundary for the DNN at $.5$ and for the CNF in case (ii) at $p^{(s)}_\theta(\mathbf{x})>10^{-11}$. In each panel, the range of $\phi$ used for training is the entire range of $\phi$ in the plot: $0 \leq \phi \leq \pi/2$ for the left panel and $-\pi/5 \leq \phi \leq \pi/5$ for the center panel. We see in Fig.~\ref{fig:Accvstheta} that case (i) (blue) matches the Bayes-optimal prediction well, as does case (ii) (maroon) for the CNF. Case (iv) (yellow) matches the DNN with CNF features for large nuisance parameters (left panel), but interestingly, not for small $\phi$ (center panel). Regardless, in all nuisance parameter scenarios we studied, the simple DNN classifier performs best at small values of the nuisance parameter, while the DNN classifier with CNF features surpasses all the other classifiers at larger nuisance parameters.

\section{Performance on the HiggsML Uncertainty Challenge Dataset}
\label{sec:challenge}

In this section we validate the performance of our DNN classifier with CNF features on the HiggsML Uncertainty Challenge Dataset~\cite{bhimji2024fairuniversehiggsmluncertainty}, designed to resemble a true uncertainty quantification task in high-energy physics. 

\subsection{Dataset and challenge description}

\begin{figure}[t]
    \centering
    \includegraphics[width=0.39\linewidth]{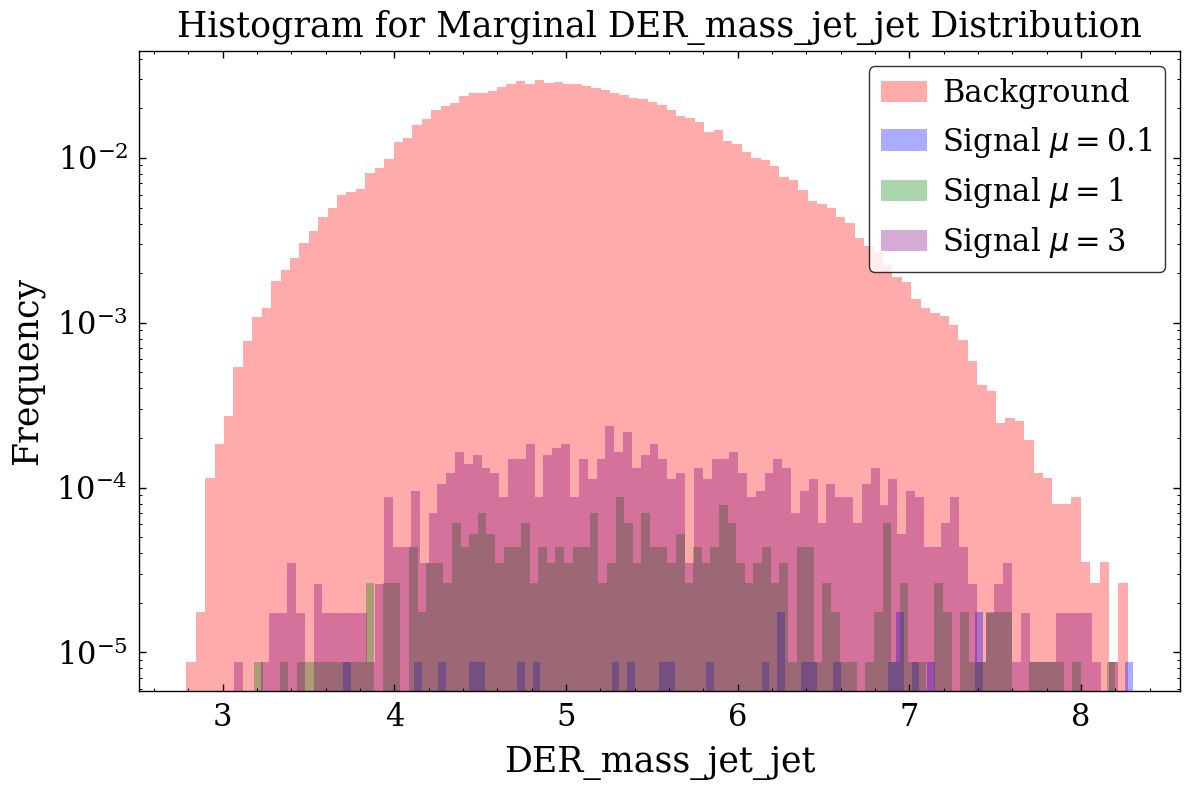}
    \includegraphics[width=0.35\linewidth]{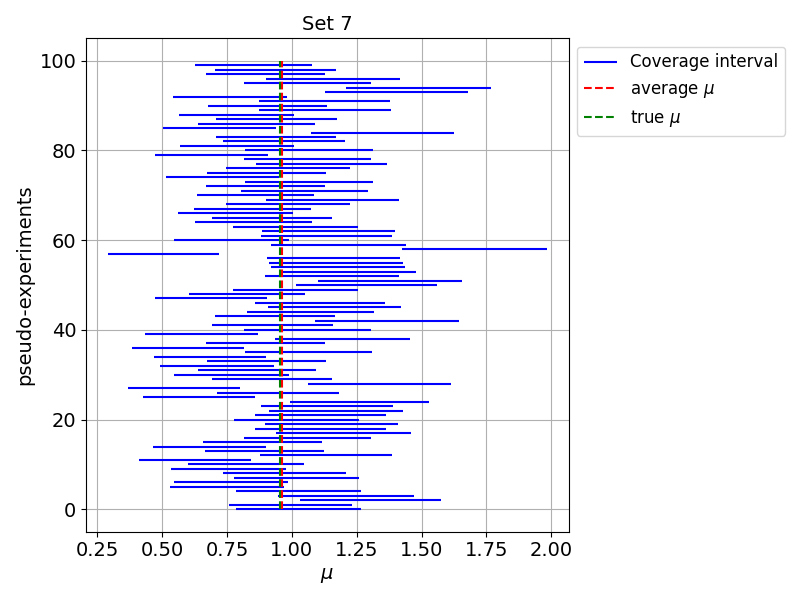}
  \caption{Setup of the HiggsML Uncertainty Challenge. The left panel shows an example of the marginal distribution of one of the 28 features (note the log scale, demonstrating a signal fraction $f_s \ll 1$), and the right panel shows an example coverage plot.}
  \label{fig:HiggsMLSetup}
\end{figure}

We briefly summarize the setup and challenge here, with more details in Appendix \ref{app:higgschal}. The data consists of simulated high-energy proton-proton particle collision events at the Large Hadron Collider (LHC). Events originating from Higgs boson decays are labeled as signal, and events generated by a diverse set of non-Higgs processes are labeled as background. The goal of the challenge is to estimate the ``signal strength parameter'' $\mu \propto f_s$ in a dataset which is a mixture of signal and background events with $\mu \in [0.1, 3]$, where $\mu = 1$ corresponds to $f_s \sim 10^{-3}$ so that there is typically a large class imbalance in the data. The data consists of 28 features $F_i$ which are various functions of particle momenta. Most of the 1D marginal distributions have poor discrimination between signal and background (see Fig.~\ref{fig:HiggsMLSetup}, left for an example); the optimal discriminating observable likely exploits the full high-dimensional structure of $p_s$ and $p_b$, making this problem an excellent test case for ML techniques. The data may be distorted or ``biased'' by 6 nuisance parameters $\{\nu_i\}$ selected to emulate realistic experimental conditions such as mis-measurement or mis-calibration. Participants can generate training data corresponding to particular choices of $\nu_i$, but do not have access to the $\nu_i$ used to generate the test data. The challenge is to correctly estimate $\mu$ along with a $1\sigma$ (68.27\%) confidence interval, such that in $N$ different pseudo-experiments each generated with different $\{\nu_i\}$, the confidence interval will cover the true value of $\mu$ in $0.6827 N$ pseudo-experiments. An example ``coverage plot'' for $N = 100$ is shown in Fig.~\ref{fig:HiggsMLSetup}, right. 

\subsection{Method overview}
\label{sec:method_overview}
\begin{figure}[t]
    \centering
    \includegraphics[width=1\linewidth]{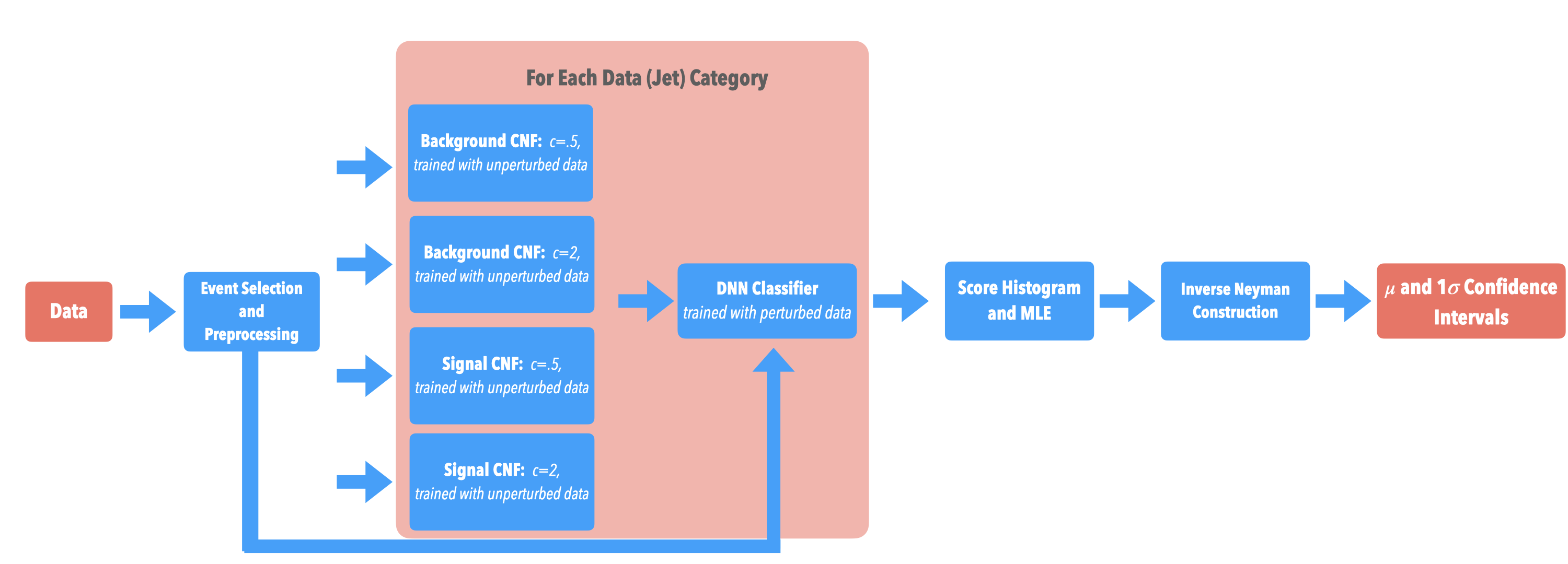}
    \caption{Flow chart for our CNF-based method of estimating $\mu$ and 1$\sigma$ confidence intervals.}
    \label{fig:methodflow}
    \vspace{-0.5cm}
\end{figure}

Our approach, schematically illustrated in Fig.~\ref{fig:methodflow}, proceeds as follows:

\paragraph{1. Event selection and preprocessing.}
We partition the dataset $\mathcal{D}$ into three categories of events depending on the jet multiplicity. Jets are sprays of collimated particles that accompany LHC events but in this case are not directly related to the signal process; empirically, we find that 0-jet events are detrimental to accurate $\mu$ estimation, but 1- and 2-jet events can change the proportion of background sub-processes in a useful way. We replace $F \to \log F$ for any feature $F$ with a marginal distribution that peaks toward zero and we then standardize all 28 features across $\mathcal{D}$ by rescaling them so they have zero mean and unit variance. The resulting feature vector $\mathbf{x} \in \mathbb{R}^{28}$ describes each event.

\begin{figure}[t]
  \centering
  \begin{subfigure}[t]{0.29\linewidth}
      \centering
    \includegraphics[width=\linewidth]{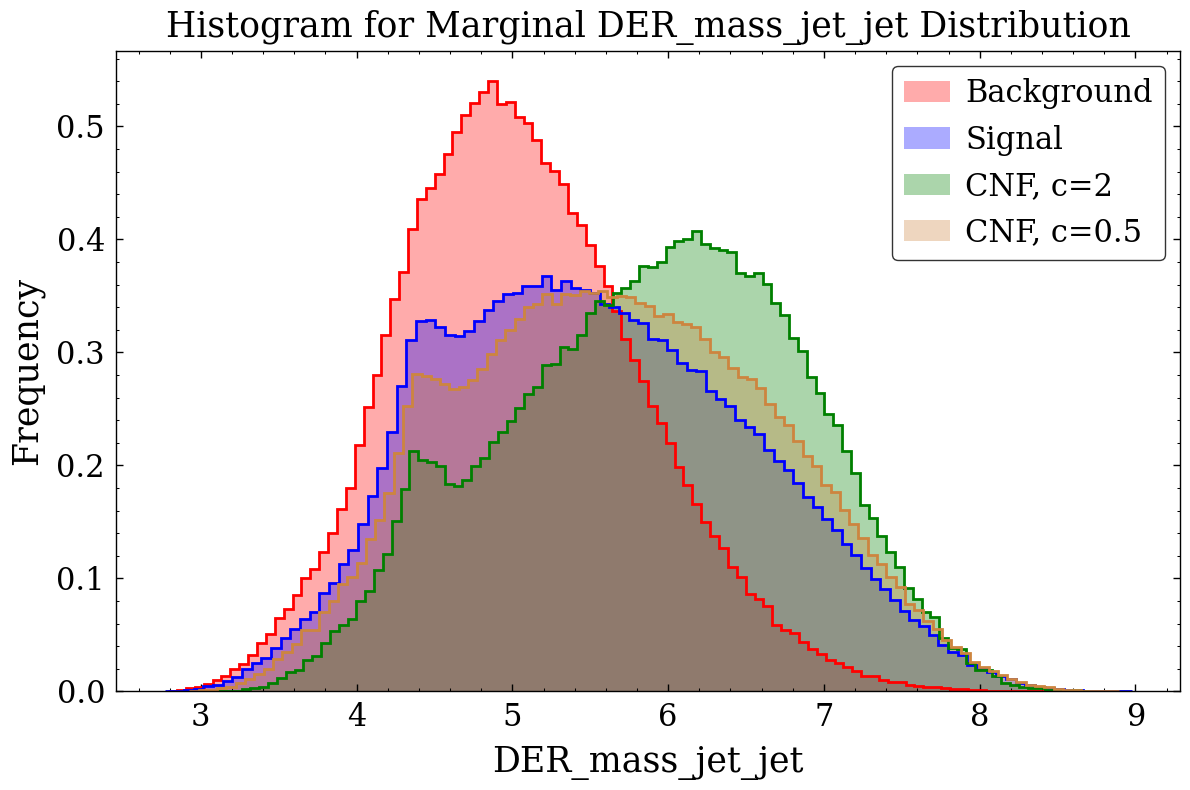}
    \label{fig:marginal_varyc}
    \end{subfigure}
\begin{subfigure}[t]{0.22\linewidth}
    \centering
    \includegraphics[width=\linewidth]{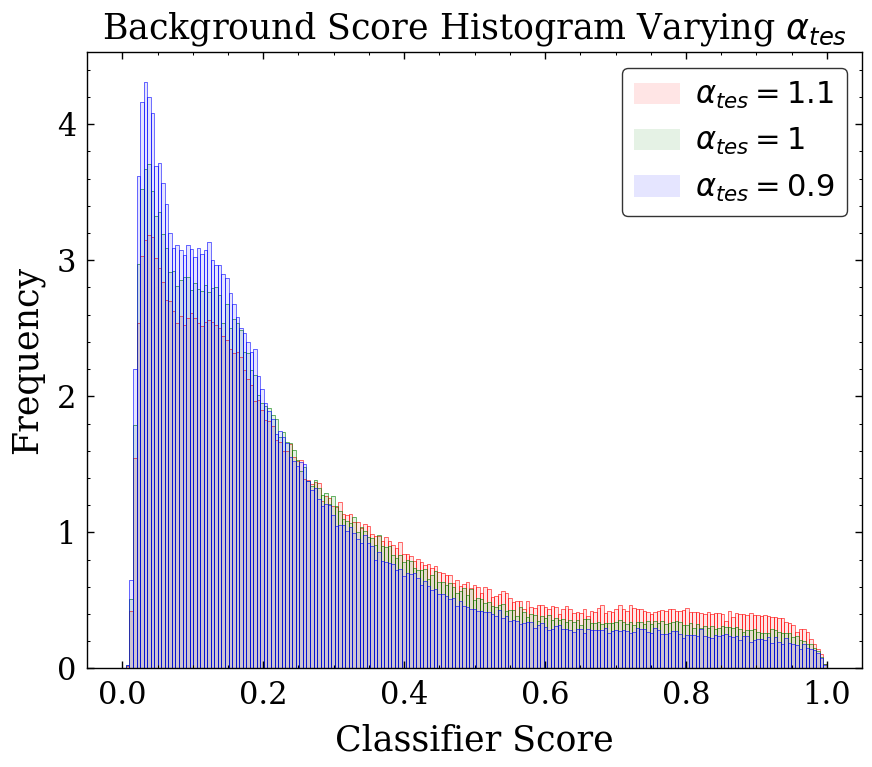}
    \label{fig:background_score}
  \end{subfigure}%
\begin{subfigure}[t]{0.22\linewidth}
    \centering
    \includegraphics[width=\linewidth]{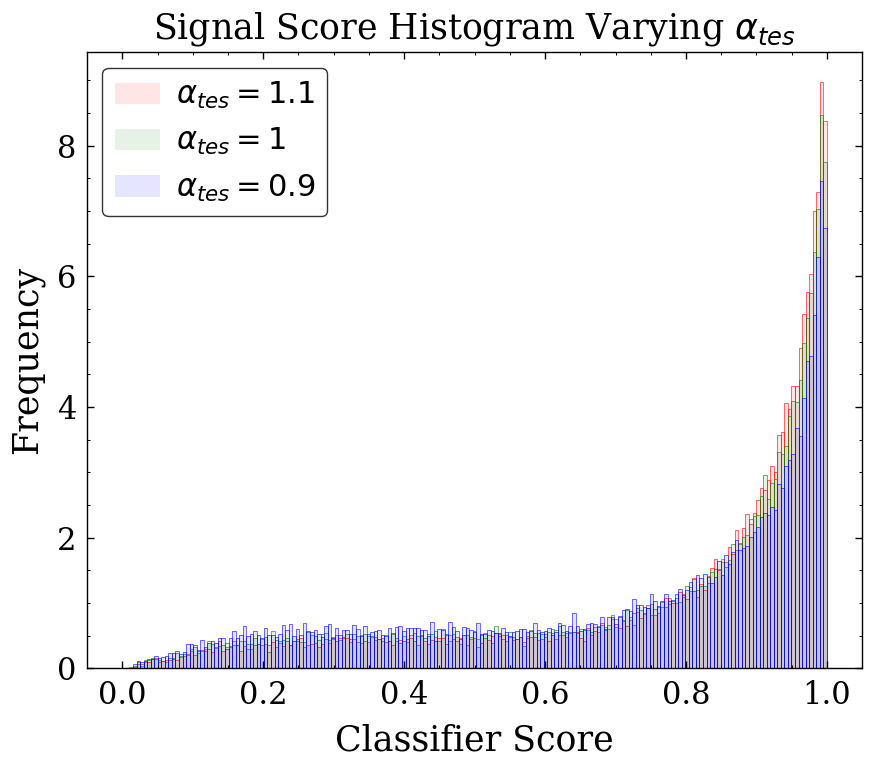}
    \label{fig:signal_score}
  \end{subfigure}
  \begin{subfigure}[t]{0.22\linewidth}
    \centering
    \includegraphics[width=\linewidth]{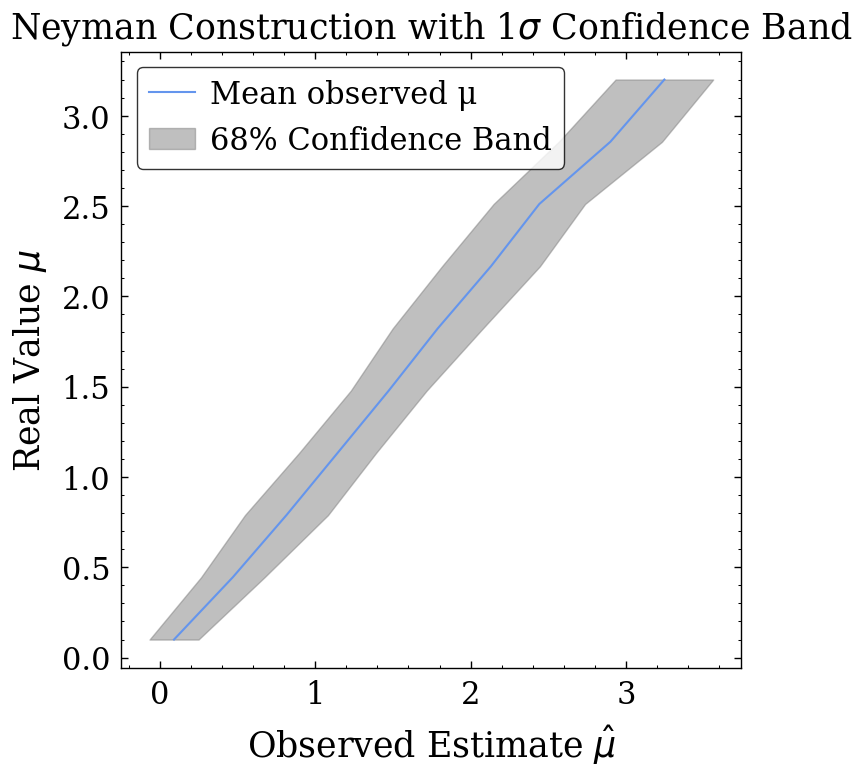}
    \label{fig:neyman}
  \end{subfigure}%
  \caption{Illustration of the CNF+DNN method for uncertainty-aware parameter estimation. The left panel plots normalized CNF distributions for the feature shown in Fig.~\ref{fig:HiggsMLSetup} for various $c$, the two center panels show DNN score histograms for signal and background varying the nuisance parameter $\alpha_{\rm tes}$, and the right panel shows the Neyman confidence belt.}
  \label{fig:HiggsMLmethod}
  \vspace{-0.3cm}
\end{figure}

\paragraph{2. CNF training.} 
Next, we train CNFs on a training set from each jet category consisting of an equal mixture of signal and background with all $\nu_i$ set to their ``unperturbed'' values (0 for additive distortions and 1 for multiplicative distortions, see Tab.~\ref{tab:nuisance-params} in App.~\ref{app:higgschal}). We extract $p_\theta^{(s)}$ and $p_\theta^{(b)}$ from CNFs with $c = 2$ and $c = 0.5$ for 1-jet and 2-jet events, for a total of 8 flows. The effect of different values of $c$ on the learned distribution can be visualized in Fig.~\ref{fig:HiggsMLmethod} (left) for $p_\theta^{(s)}$. As $c$ increases, the CNF picks out more signal-rich regions. Due to this fact, we chose one CNF with $c>1$ to be more robust to nuisance parameters, and another with $c<1$ to capture more of the total signal density.

\paragraph{3. DNN training.} We create an augmented training set composed of 1000 sub-datasets, each generated using a different combination of $\nu_i$ and with an equal fraction of signal and background. For each event $\mathbf{x}$, we then compute the discrimination functions $\Phi^{(s,b)}(\mathbf{x})= \frac{p_\theta^{(s,b)}(\mathbf{x})}{1+p_\theta^{(s,b)}(\mathbf{x})}$ and append them to the list of features $\mathbf{x}$. A two-headed DNN classifier (one head per jet category) trained with binary cross-entropy loss then uses both the original kinematic features $\mathbf{x}$ and the CNF-based discriminating functions $\Phi$ to differentiate signal from background.

\paragraph{4. Neyman construction for confidence intervals.}
After training, we create two sets of fine-binned histograms (one for signal events and one for background events) of the DNN classifier scores $r$, where each set is composed of histograms corresponding to a dense grid of values of the two nuisance parameters $\alpha_{\rm jes}$, and $\alpha_{\rm tes}$ which we empirically find affect the signal/background discrimination the most. We show the effect of varying $\alpha_{\rm tes}$ on the background and signal classifier scores in Fig.~\ref{fig:HiggsMLmethod}; note that $\alpha_{\rm tes} > 1$ pushes background probability mass toward a classifier score of 1, indicating mis-classification of background as signal, but the signal classifier is much more robust to variations in $\alpha_{\rm tes}$. Using these histograms, we build, with spline interpolation, a binned likelihood function $\mathcal{L}(\{r_k\}|\mu,\alpha_{\rm jes},\alpha_{\rm tes})$ that depends on $\mu$ and the two nuisance parameters $\alpha_{\rm jes}$ and $\alpha_{\rm tes}$. We can thus obtain point estimates $\hat{\mu}$, $\hat{\alpha}_{\rm jes}$, and $\hat{\alpha}_{\rm tes}$ by likelihood maximization on a set of data with classifier scores $\{r_k\}$.

Using the point estimates $\hat{\mu}$, we build a confidence belt (Fig.~\ref{fig:HiggsMLmethod}, right) using the Neyman construction~\cite{neyman1937outline} by computing the distribution of $\hat{\mu}$ for various values of real $\mu$ given many test sets generated using random draws of all six nuisance parameters. This construction implicitly profiles over the nuisance parameters, and inverting the confidence belt provides a $1\sigma$ confidence interval of $\mu$ given $\hat{\mu}$ that should cover the true $\mu$~ 68.27\% of the time. 

Given a test set, we evaluate the pre-trained CNF discrimination functions on the set, append these to the data, and pass the test data through the pre-trained DNN to build a histogram of scores $\{r_k\}$. We then determine the $\hat{\mu}$ that maximizes the binned likelihood function $\mathcal{L}(\{r_k\}|\mu,\alpha_{\rm jes},\alpha_{\rm tes})$, and then extract the confidence interval at that value of $\hat{\mu}$ from the previously-constructed Neyman bands.

\subsection{Results}
The performance of our method is summarized in Tab.~\ref{tab:results}. On a test set comprised of $10\times100$ pseudo-experiments, with 10 values of $\mu$ and 100 datasets of varying $\nu_i$ per value of $\mu$, we compare 1) our method, 2) a purely likelihood based method (common in HEP analyses) to learn Eq.~(\ref{eq:nophilikli}) described in detail in App. \ref{app:np_method_overview}, and 3) a baseline method using a DNN classifier trained on perturbed data and MLE for $\mu$ only. All three procedures employ a Neyman construction for confidence intervals (CI). The reported metrics are a score defined in App. \ref{app:higgschal} (log scale; larger is better), the average size of the confidence interval (smaller is better), the coverage fraction, i.e the fraction of events where $\mu$ resides in the predicted CI (for $1\sigma$ this should be close to .6827), the root-mean-square error (RMSE) of $\hat{\mu}$ vs $\mu$, and the run time of each method on a 4xA100 node at the Perlmutter supercomputer \cite{nersc-perlmutter-architecture}. Our method achieves the best performance among the three strategies. 

\begin{table}[h!]
  \caption{Comparison of the three methods for $10\times100$ pseudo-experiments.}
  \label{tab:results}
  \vskip 0.1in
  \begin{center}
    \small
    \setlength{\tabcolsep}{5pt}
    \begin{tabular}{lccccc}
      \toprule
      \textbf{Method} & \textbf{Score} & \textbf{Interval} & \textbf{Coverage Fraction} & \textbf{RMSE} & \textbf{Runtime (min)}\\
      \midrule
      CNF+DNN (this paper) & \textbf{0.823} & \textbf{0.438} & \textbf{0.672} & \textbf{0.191} & 5.0\\
      MLE with $\nu_i$ & 0.001 & 0.998 & 0.701 & 0.575 & 41.0 \\
      Baseline MLE for $\mu$ only& -11.152 & 1.34 & 0.431 & 1.575 & \textbf{3.0} \\
      \bottomrule
    \end{tabular}
  \end{center}
  \vskip -0.1in
\end{table}

We also report for our method the metrics with a $100 \times 100$ (i.e $100$ values of $\mu$) pseudo-experiments bootstrapped 1000 times for precision. In this case, we achieve a coverage of \textbf{0.6689} and an interval length of \textbf{0.4947}, only marginally different than above \cite{bhimji2024fairuniversehiggsmluncertainty}. These results emphasize the importance of CNF features; with them, we can achieve half the CI size with the same coverage and a runtime $\sim10\times$ faster as compared to a likelihood-based method. We emphasize that our method is capable of picking out the $100-3000$ signal events among a dataset of $10^{6}$ events, with such a precision that the $1\sigma$ error bar spans just $150-500$ events, despite the degenerate effects of systematic errors. 

\section{Conclusion}
We have introduced a novel method for estimating model parameters proportional to rare signal fractions in the presence of systematic uncertainties. Our method combines contrastive normalizing flows (CNFs) with a DNN classifier trained on features derived from the CNF. Our method shows robustness to nuisance‐parameter–induced deformations compared to traditional classifiers, which we have illustrated on a toy example of Gaussian data. We further demonstrated the performance of our method in the HiggsML Uncertainty Challenge, where we obtain calibrated frequentist confidence intervals on the signal strength that remain reliable even under significant systematics/domain shifts. In addition to future studies on the generative capacity of a CNF, we leave the task of finding an optimal $c$ for a parameter estimation problem for future work. We also plan to investigate the physical interpretation of the learned features in either the CNFs or DNN for the Higgs decay process underlying the challenge, for which analytic distributions may be calculated using quantum field theory.
\begin{ack}
We thank Ben Hooberman and Joshua Foster for enlightening discussions. IE especially thanks the organizers of the HiggsML Uncertainty Challenge --- Wahid Bhimji, Ragansu Chakkappai, Sascha Diefenbacher, David Rousseau, Benjamin Nachman, Shih-Chieh Hsu, Po-Wen Chang, Chris Harris, Ihsan Ullah, and Yulei Zhang --- for the organization of the challenge and workshops, many fruitful conversations, and ample technical support.  
This material is based upon work supported by the U.S. Department of Energy, Office of Science, Office of High Energy Physics, under Award Number DE-SC0023704. This work used the \texttt{TAMU FASTER} cluster at Texas A\&M University through allocation 240449 from the Advanced Cyberinfrastructure Coordination Ecosystem: Services \& Support (ACCESS) program~\cite{10.1145/3569951.3597559}, which is supported by National Science Foundation grants \#2138259, \#2138286, \#2138307, \#2137603, and \#213829.
\end{ack}

\bibliographystyle{JHEP.bst}
\bibliography{biblo}

\providecommand{\href}[2]{#2}\begingroup\raggedright\begin{thebibliography}{10}

\bibitem{bhimji2024fairuniversehiggsmluncertainty}
W.~Bhimji, P.~Calafiura, R.~Chakkappai, P.-W. Chang, Y.-T. Chou,
  S.~Diefenbacher et~al., \emph{{FAIR Universe HiggsML Uncertainty Challenge
  Competition}},  \href{https://arxiv.org/abs/2410.02867}{{\tt 2410.02867}}.

\bibitem{louppe2020adversarialvariationaloptimizationnondifferentiable}
G.~Louppe, J.~Hermans and K.~Cranmer, \emph{Adversarial variational
  optimization of non-differentiable simulators},
  \href{https://arxiv.org/abs/1707.07113}{{\tt 1707.07113}}.

\bibitem{ovadia2019trustmodelsuncertaintyevaluating}
Y.~Ovadia, E.~Fertig, J.~Ren, Z.~Nado, D.~Sculley, S.~Nowozin et~al.,
  \emph{{Can You Trust Your Model's Uncertainty? Evaluating Predictive
  Uncertainty Under Dataset Shift}},
  \href{https://arxiv.org/abs/1906.02530}{{\tt 1906.02530}}.

\bibitem{pmlr-v139-koh21a}
P.~W. Koh, S.~Sagawa, H.~Marklund, S.~M. Xie, M.~Zhang, A.~Balsubramani et~al.,
  \emph{Wilds: A benchmark of in-the-wild distribution shifts},  in
  \emph{Proceedings of the 38th International Conference on Machine Learning}
  (M.~Meila and T.~Zhang, eds.), vol.~139 of \emph{Proceedings of Machine
  Learning Research}, pp.~5637--5664, PMLR, 18--24 Jul, 2021.

\bibitem{arjovsky2020invariantriskminimization}
M.~Arjovsky, L.~Bottou, I.~Gulrajani and D.~Lopez-Paz, \emph{Invariant risk
  minimization},  \href{https://arxiv.org/abs/1907.02893}{{\tt 1907.02893}}.

\bibitem{sagawa2020distributionallyrobustneuralnetworks}
S.~Sagawa, P.~W. Koh, T.~B. Hashimoto and P.~Liang, \emph{Distributionally
  robust neural networks for group shifts: On the importance of regularization
  for worst-case generalization},  \href{https://arxiv.org/abs/1911.08731}{{\tt
  1911.08731}}.

\bibitem{malinin2018predictiveuncertaintyestimationprior}
A.~Malinin and M.~Gales, \emph{Predictive uncertainty estimation via prior
  networks},  \href{https://arxiv.org/abs/1802.10501}{{\tt 1802.10501}}.

\bibitem{lipton2017mythosmodelinterpretability}
Z.~C. Lipton, \emph{The mythos of model interpretability},
  \href{https://arxiv.org/abs/1606.03490}{{\tt 1606.03490}}.

\bibitem{Baldi_2016}
P.~Baldi, K.~Cranmer, T.~Faucett, P.~Sadowski and D.~Whiteson,
  \emph{Parameterized neural networks for high-energy physics},
  \href{http://dx.doi.org/10.1140/epjc/s10052-016-4099-4}{\emph{The European
  Physical Journal C} {\bf 76} (Apr., 2016) }.

\bibitem{louppe2017learningpivotadversarialnetworks}
G.~Louppe, M.~Kagan and K.~Cranmer, \emph{Learning to pivot with adversarial
  networks},  \href{https://arxiv.org/abs/1611.01046}{{\tt 1611.01046}}.

\bibitem{Ghosh_2021}
A.~Ghosh, B.~Nachman and D.~Whiteson, \emph{Uncertainty-aware machine learning
  for high energy physics},
  \href{http://dx.doi.org/10.1103/physrevd.104.056026}{\emph{Physical Review D}
  {\bf 104} (Sept., 2021) }.

\bibitem{dorigo2022endtoendoptimizationparticlephysics}
T.~Dorigo, A.~Giammanco, P.~Vischia, M.~Aehle, M.~Bawaj, A.~Boldyrev et~al.,
  \emph{Toward the end-to-end optimization of particle physics instruments with
  differentiable programming: a white paper},
  \href{https://arxiv.org/abs/2203.13818}{{\tt 2203.13818}}.

\bibitem{dinh2017densityestimationusingreal}
L.~Dinh, J.~Sohl-Dickstein and S.~Bengio, \emph{{Density estimation using Real
  NVP}},  \href{https://arxiv.org/abs/1605.08803}{{\tt 1605.08803}}.

\bibitem{kingma2018glowgenerativeflowinvertible}
D.~P. Kingma and P.~Dhariwal, \emph{{Glow: Generative Flow with Invertible $1
  \times 1$ Convolutions}},  \href{https://arxiv.org/abs/1807.03039}{{\tt
  1807.03039}}.

\bibitem{papamakarios2021normalizingflowsprobabilisticmodeling}
G.~Papamakarios, E.~Nalisnick, D.~J. Rezende, S.~Mohamed and
  B.~Lakshminarayanan, \emph{Normalizing flows for probabilistic modeling and
  inference},  \href{https://arxiv.org/abs/1912.02762}{{\tt 1912.02762}}.

\bibitem{nalisnick2019detectingoutofdistributioninputsdeep}
E.~Nalisnick, A.~Matsukawa, Y.~W. Teh and B.~Lakshminarayanan, \emph{Detecting
  out-of-distribution inputs to deep generative models using typicality},
  \href{https://arxiv.org/abs/1906.02994}{{\tt 1906.02994}}.

\bibitem{ren2019likelihoodratiosoutofdistributiondetection}
J.~Ren, P.~J. Liu, E.~Fertig, J.~Snoek, R.~Poplin, M.~A. DePristo et~al.,
  \emph{Likelihood ratios for out-of-distribution detection},
  \href{https://arxiv.org/abs/1906.02845}{{\tt 1906.02845}}.

\bibitem{schmier2023positivedifferencedistributionimage}
R.~Schmier, U.~Köthe and C.-N. Straehle, \emph{Positive difference
  distribution for image outlier detection using normalizing flows and
  contrastive data},  \href{https://arxiv.org/abs/2208.14024}{{\tt
  2208.14024}}.

\bibitem{9879060}
S.~Cao and Z.~Zhang, \emph{Deep hybrid models for out-of-distribution
  detection},  in \emph{2022 IEEE/CVF Conference on Computer Vision and Pattern
  Recognition (CVPR)}, pp.~4723--4733, 2022.
\newblock \href{http://dx.doi.org/10.1109/CVPR52688.2022.00469}{DOI}.

\bibitem{Andreassen_2020}
A.~Andreassen and B.~Nachman, \emph{Neural networks for full phase-space
  reweighting and parameter tuning},
  \href{http://dx.doi.org/10.1103/physrevd.101.091901}{\emph{Physical Review D}
  {\bf 101} (May, 2020) }.

\bibitem{cranmer2016approximatinglikelihoodratioscalibrated}
K.~Cranmer, J.~Pavez and G.~Louppe, \emph{Approximating likelihood ratios with
  calibrated discriminative classifiers},
  \href{https://arxiv.org/abs/1506.02169}{{\tt 1506.02169}}.

\bibitem{neyman1937outline}
J.~Neyman, \emph{Outline of a theory of statistical estimation based on the
  classical theory of probability}, {\emph{Philosophical Transactions of the
  Royal Society of London. Series A, Mathematical and Physical Sciences} {\bf
  236} (1937) 333--380}.

\bibitem{nersc-perlmutter-architecture}
{National Energy Research Scientific Computing Center}, ``{Perlmutter} system
  architecture.''
  \url{https://docs.nersc.gov/systems/perlmutter/architecture/}.

\bibitem{10.1145/3569951.3597559}
T.~J. Boerner, S.~Deems, T.~R. Furlani, S.~L. Knuth and J.~Towns,
  \emph{{ACCESS: Advancing Innovation: NSF’s Advanced Cyberinfrastructure
  Coordination Ecosystem: Services \& Support}},  in \emph{Practice and
  Experience in Advanced Research Computing 2023: Computing for the Common
  Good}, PEARC '23, (New York, NY, USA), p.~173–176, Association for
  Computing Machinery, 2023.
\newblock \href{http://dx.doi.org/10.1145/3569951.3597559}{DOI}.

\end{thebibliography}\endgroup

\medskip


\appendix

\section{Architecture and Training of a CNF}
\label{app:ArchCNF}

Training a Contrastive Normalizing Flow (CNF) builds on a standard normalizing flow (NF) architecture, which itself is composed of a sequence of invertible coupling transformations. Below, we first review the structure of a typical NF with affine coupling layers, then describe the modifications that are required to train a CNF.

\subsection{Normalizing flow with coupling layers}
A normalizing flow learns a target density $p_X(\mathbf{x})$ by applying an invertible mapping $f_\theta$ to a simple base distribution $p_z(\mathbf{z})$, typically a standard Gaussian. The mapping is defined as
\begin{equation}
      \mathbf{x} = f_\theta(\mathbf{z}) = f_L \circ f_{L-1} \circ \cdots \circ f_1(\mathbf{z}),
\end{equation}
where each coupling layer $f_i$ is chosen to admit a tractable Jacobian determinant. In the popular affine coupling scheme \cite{dinh2017densityestimationusingreal}, each layer partitions its input $\mathbf{u}\in\mathbb{R}^d$ into two complementary subsets $\mathbf{u} = (\mathbf{u}_a,\mathbf{u}_b)$ via a binary mask $M\in\{0,1\}^d$ such that $\mathbf{u}_a = M \odot \mathbf{u}, \quad \mathbf{u}_b = (1-M) \odot \mathbf{u}$. The affine coupling transformation defines
\begin{equation}
\label{eq:flowcoupling}
\mathbf{y}_a = \mathbf{u}_a \odot \exp\bigl(s(\mathbf{u}_b))\bigr) + t(\mathbf{u}_b), \qquad
\mathbf{y}_b = \mathbf{u}_b,
\end{equation}
where $s(\mathbf{u}_b)$ (the scale) and $t(\mathbf{u}_b)$ (the translation) are outputs of a small neural network (in our case a DNN) given $\mathbf{u}_b$ as input. Masks $M$ are alternated between layers so that each dimension is transformed. The transformed variables $(\mathbf{y}_a,\mathbf{y}_b)$ denote, respectively, the updated subset and the pass-through (identity) subset at each layer. 

Our DNN architecture has 5 hidden layers, 64 neurons per layer, and GELU activations. All training uses the Adam optimizer with a learning rate of $1\times10^{-3}$ and default $\beta$ coefficients $(0.9,0.999)$.

\subsection{Modifications for contrastive normalizing flow}

\begin{figure}[t!]
  \centering
  \begin{subfigure}[t]{0.5\linewidth}
    \centering
    \includegraphics[width=\linewidth]{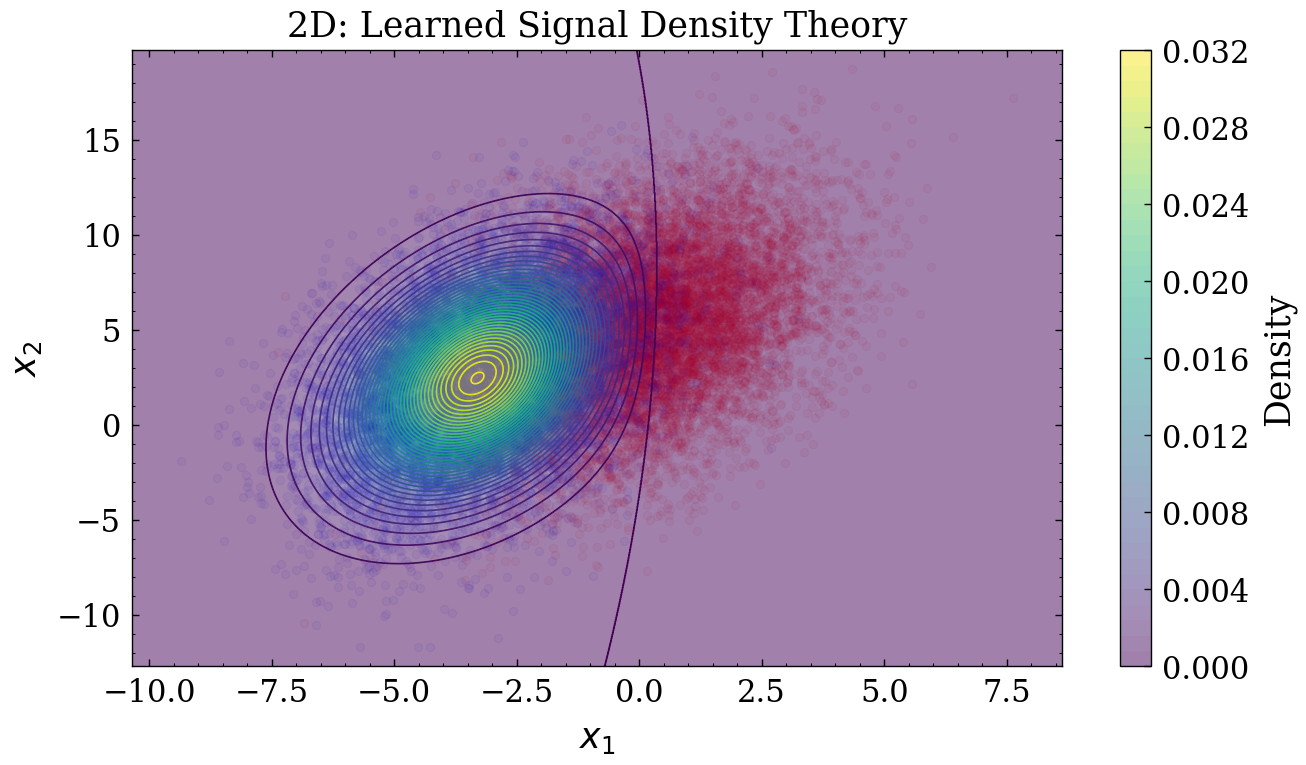}
    \caption{$c=0.1$}
  \end{subfigure}%
  \begin{subfigure}[t]{0.5\linewidth}
    \centering
    \includegraphics[width=\linewidth]{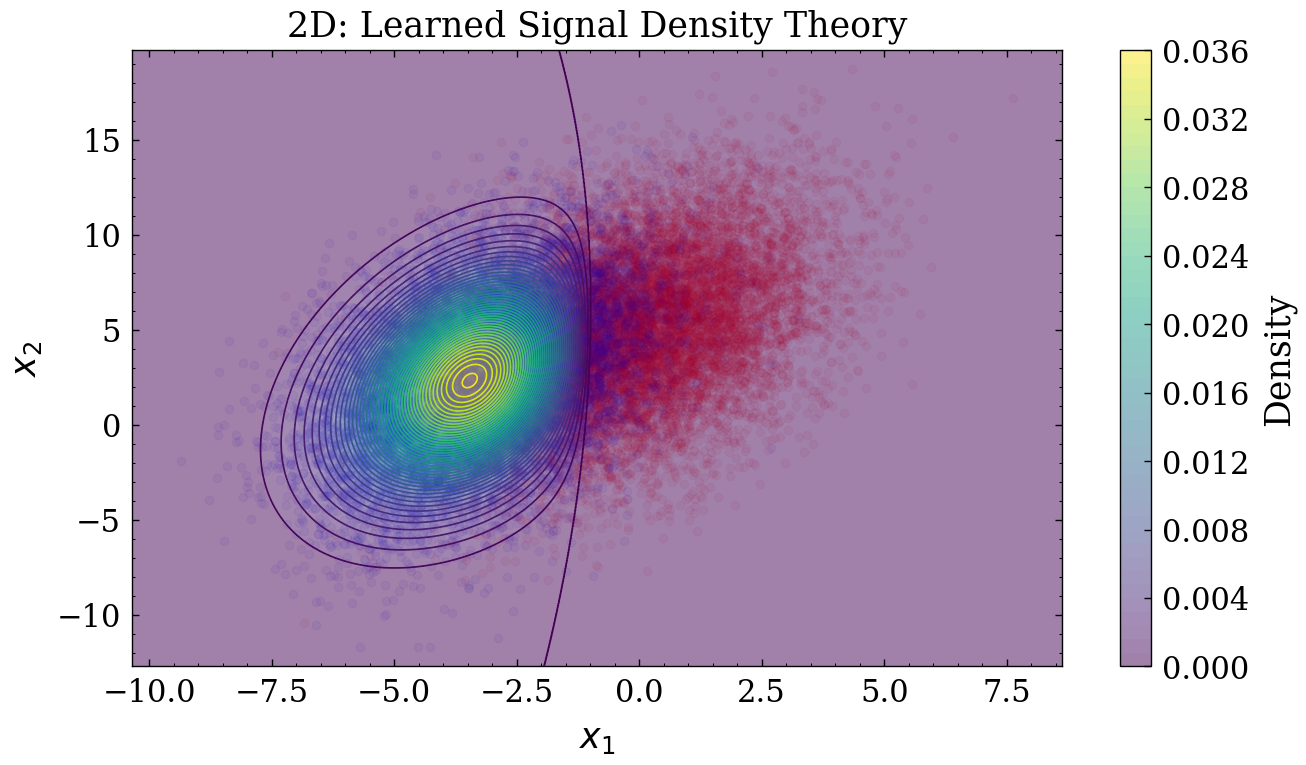}
    \caption{$c=0.5$}
  \end{subfigure}%
  \\[1ex] 
  \begin{subfigure}[t]{0.5\linewidth}
    \centering
    \includegraphics[width=\linewidth]{images/theory_c1_thick_few.png}
    \caption{$c=1$}
  \end{subfigure}%
  \begin{subfigure}[t]{0.5\linewidth}
    \centering
    \includegraphics[width=\linewidth]{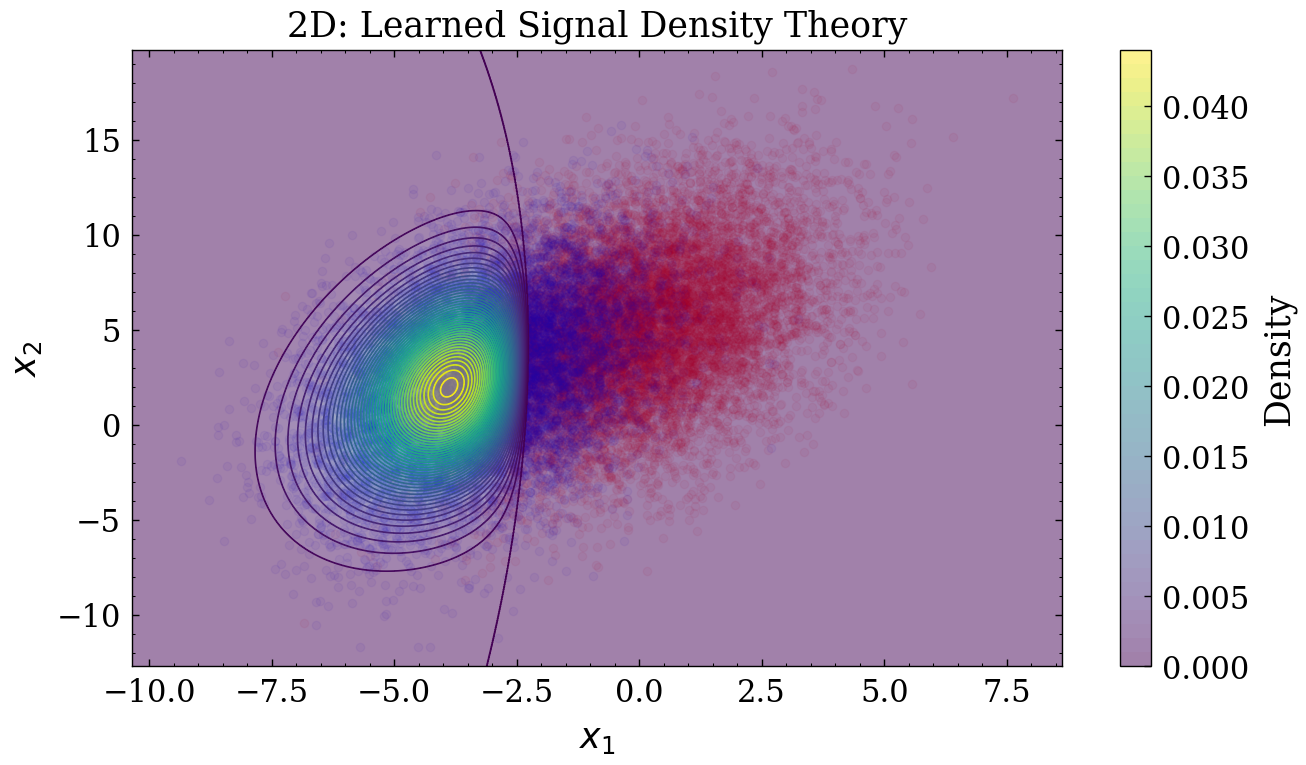}
    \caption{$c=2$}
  \end{subfigure}%
  \\[1ex]
  \begin{subfigure}[t]{0.5\linewidth}
    \centering
    \includegraphics[width=\linewidth]{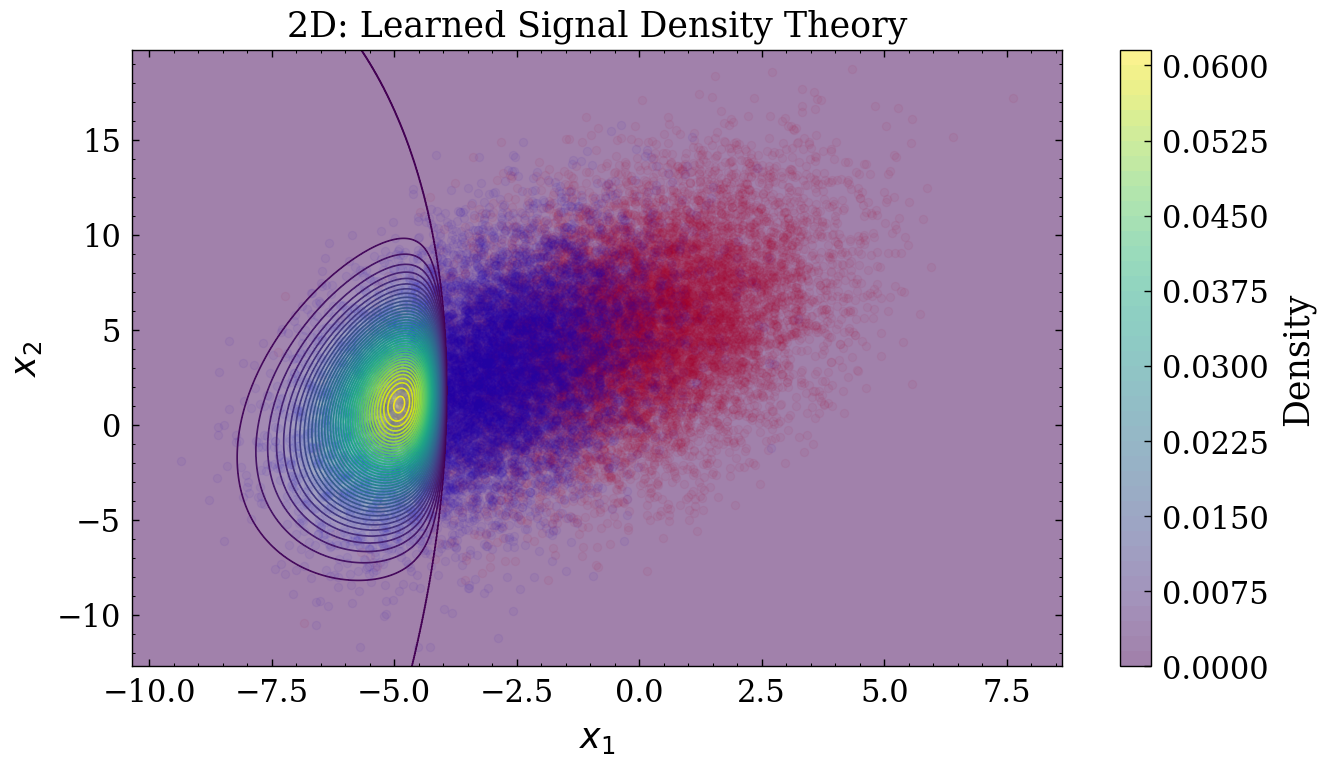}
    \caption{$c=10$}
  \end{subfigure}%
  \begin{subfigure}[t]{0.5\linewidth}
    \centering
    \includegraphics[width=\linewidth]{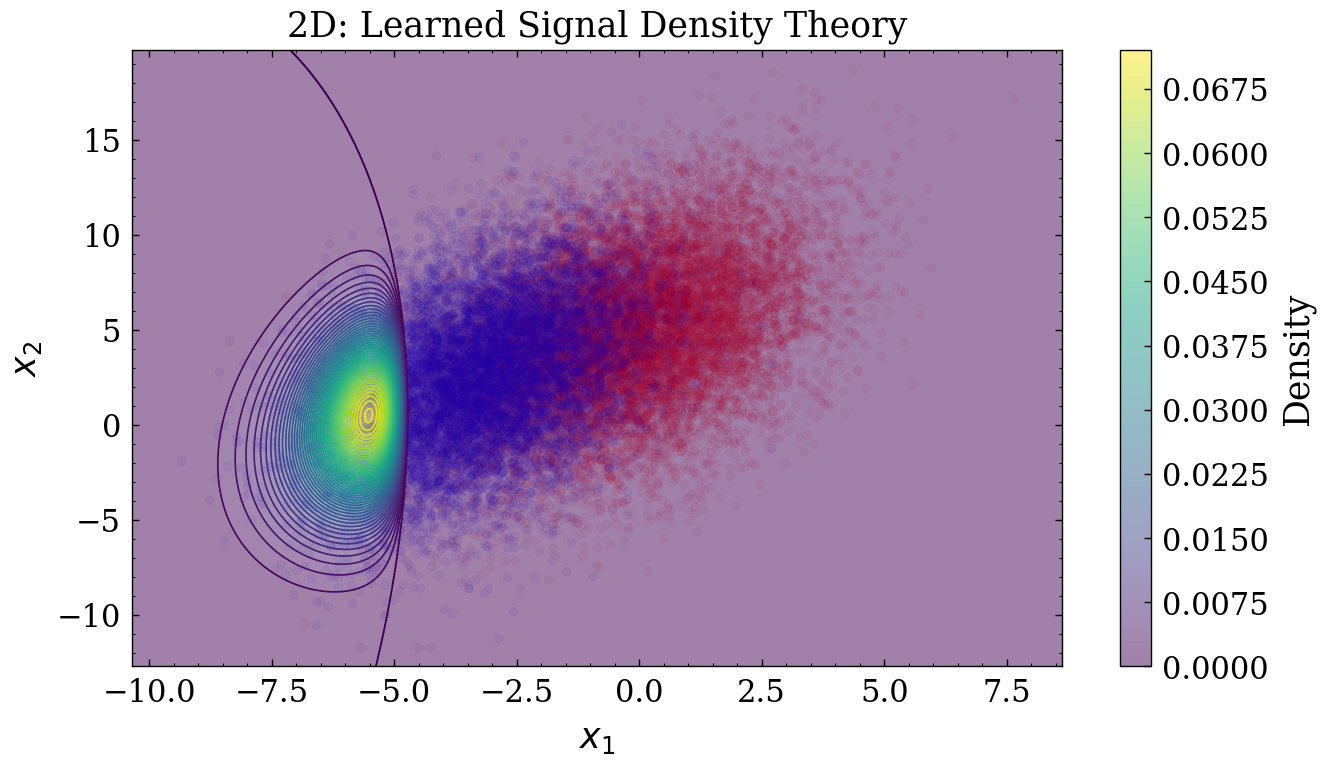}
    \caption{$c=20$}
  \end{subfigure}
  \caption{Contours of the loss-minimizing $p^{(s)}_\theta(\mathbf{x})$ from Eq.~(\ref{eq:opt}) on the toy example of 2-dimensional Gaussians from Sec.~\ref{sec:toy}, for various values of $c$.}
  \label{fig:effofc}
\end{figure}

Due to CNF's contrastive objective, training a CNF in practice is complicated by two factors. The first is the existence of a local minimum of the loss where the CNF rejects all samples by letting the contrastive term $c \, \log p^{(s)}_\theta(\mathbf{x}_b)\rightarrow-\infty$ faster than the rate that $\log p^{(s)}_\theta(\mathbf{x}_s)\rightarrow-\infty$. To prevent this mode collapse, a clamping procedure is required to bound the loss by some number $k$. That is, at each step in gradient descent, if the loss decreases below some value $k$ we set the loss to $k$. We find $k=-50$ to be a reasonable setting for various problems, as also noted in Ref.~\cite{schmier2023positivedifferencedistributionimage}. The second difficulty in is the sharp boundary that exists at $c p^{(s)}_\theta(\mathbf{x}_b)= p^{(s)}_\theta(\mathbf{x}_s)$ in the distribution the CNF attempts to approximate, Eq.~(\ref{eq:opt}). Due to this, a simple flow composed only of the affine coupling layers described in Eq. \ref{eq:flowcoupling} will have instabilities in training, as affine-coupling flows have an inductive bias toward smooth functions. An empirical solution we find ensures stable training is the introduction of an alternating coupling architecture, where even coupling layers are given by a standard affine transformation with the addition of a gating mechanism,
\begin{equation}
\mathbf{y}_a = g(\mathbf{u}_b) \odot \Bigl(\mathbf{u}_a \odot \exp\bigl(s(\mathbf{u}_b)\bigr) + t(\mathbf{u}_b)\Bigr) + \bigl(1 - g(\mathbf{u}_b)\bigr) \odot \mathbf{u}_a, \qquad
\mathbf{y}_b  = \mathbf{u}_b,
\end{equation}
where again $s(\mathbf{u}_b)$ and $t(\mathbf{u}_b)$ are scale and translation functions, and $g(\mathbf{u}_b) \in [0,1]^{d}$ is a gating function, all given as outputs of our DNN.  We apply a sigmoid to the outputs corresponding to $g_\theta$ to ensure values in $[0,1]$. For odd coupling layers, a quadratic transformation is used,
\begin{equation}
\mathbf{y}_a  = \frac{1}{2}\,\alpha(\mathbf{u}_b )\,\mathbf{u}_a ^2 + \beta(\mathbf{u}_b)\,\mathbf{u}_a  + \gamma(\mathbf{u}_b), \qquad
\mathbf{y}_b  = \mathbf{u}_b,
\end{equation}
where $\alpha(\mathbf{u}_b )$, $\beta(\mathbf{u}_b)$, and $ \gamma(\mathbf{u}_b)$ are functions parameterized by a small DNN. We stack $L=30$ coupling layers (15 gated affine and 15 quadratic, interleaved) with one DNN corresponding to each coupling layer.

\section{Varying $c$ in the Toy Example}
\label{app:vary_c}

Figure \ref{fig:effofc} is a visualization of different values of c on the toy problem of 2-dimensional Gaussians from Sec.~\ref{sec:toy}. As expected, larger values of $c$ reject more background at the expense of cutting out more of the signal.

\section{Illustrative Example of CNFs: MNIST}
\label{app:mnist}
To make the abstract discussion in Secs.~\ref{sec:toy} and \ref{sec:challenge} more concrete, we train Contrastive Normalizing Flows on MNIST. The architecture follows App.~\ref{app:ArchCNF}: each flow has $30$ affine coupling layers, the coupling-layer DNNs are depth‑$5$ MLPs with hidden dimension $64$ and GELU activations, and we train with Adam (learning rate $10^{-3}$, $\beta_{1,2}=(0.9,0.999)$) with loss clamping at $k=-50$. We also train a standard NF (with the same architecture but with no contrastive loss) for comparison. This architecture is not ideal for image generation, but our goal is simply to illustrate visually the effect of the contrastive loss. We do not attempt to make any quantitative statements in this appendix.
\subsection{Generative example: $0$ vs.~$8$}
\label{app:mnist:gen}
\begin{figure}[t!]
    \centering
    \begin{subfigure}{0.4\textwidth}
        \centering
        \includegraphics[width=\textwidth]{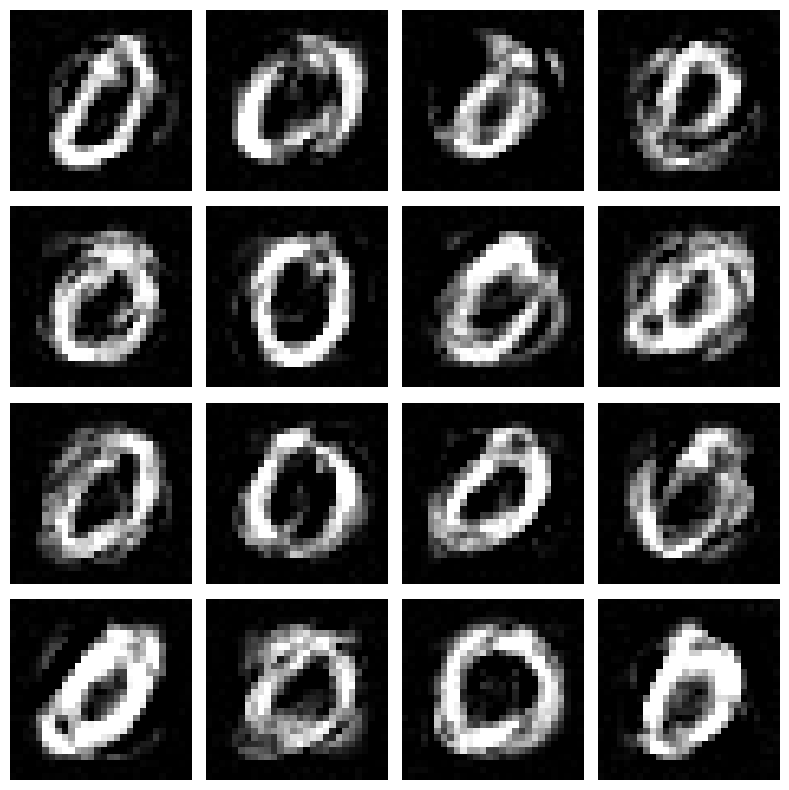}   
    \end{subfigure}
    \hspace{0.5cm}
        \centering
    \begin{subfigure}{0.4\textwidth}
        \centering
     \includegraphics[width=\textwidth]{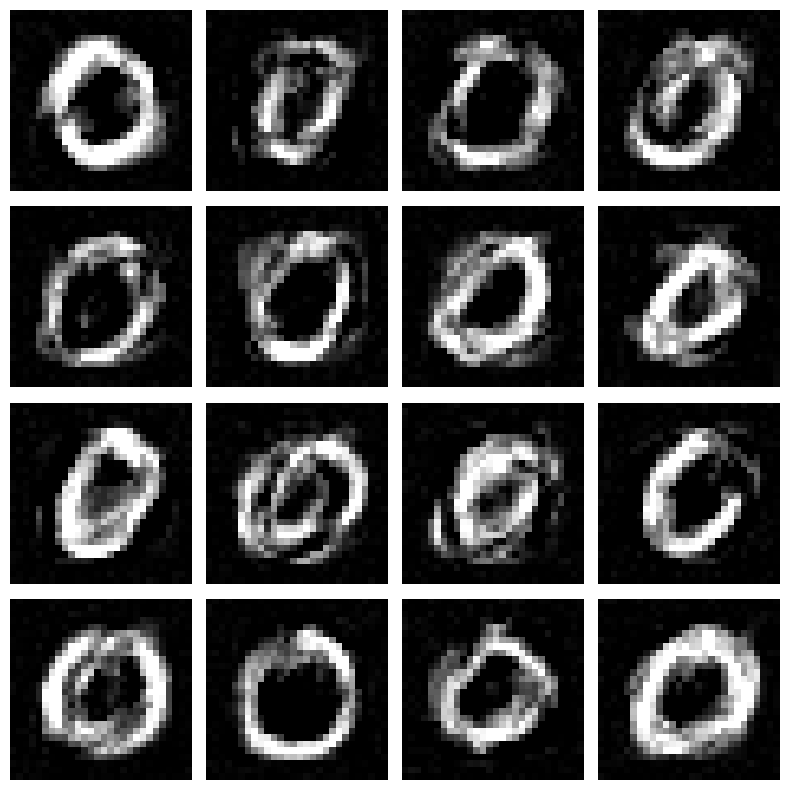}   
    \end{subfigure}
    \caption{Samples from a NF trained on just MNIST zeros (left), compared to a CNF trained with zeros ($\mathcal D_s$) in contrast to eights ($\mathcal D_b$) (right).}
    \vspace{-0.5cm}
    \label{fig:mnist_gen}
\end{figure}

We first fit a single CNF to the pair of classes \(\mathcal D_s=\{\,\text{all zeros}\,\}\) and \(\mathcal D_b=\{\,\text{all eights}\,\},\)
using $c=2$. We also train a train a NF on just \(\mathcal D_s=\{\,\text{all zeros}\,\}\) with the same architecture. After training, we draw samples from $p_{\theta}^{(s)}(\mathbf x)$  by drawing a latent $\mathbf{z}\sim\mathcal N(0,1)$ and inverting the flow. Figure \ref{fig:mnist_gen} shows a random grid of $4\times4$ samples from both the NF and the CNF. Comparing the two, visually, it seems the CNF samples are perhaps slightly sharper, and more samples seem to be round circles as opposed to oblong ovals. This hints at the possibility of the improved sample generation in specific use cases from training CNFs, which we intend to explore in future work. 
\subsection{Classification examples: $9$ vs.~$4$ and $8$ vs.~$0$}
\label{app:mnist:cls}
\begin{figure}[t!]
    \centering
    \begin{subfigure}{0.4\textwidth}
        \centering
        \includegraphics[width=\textwidth]{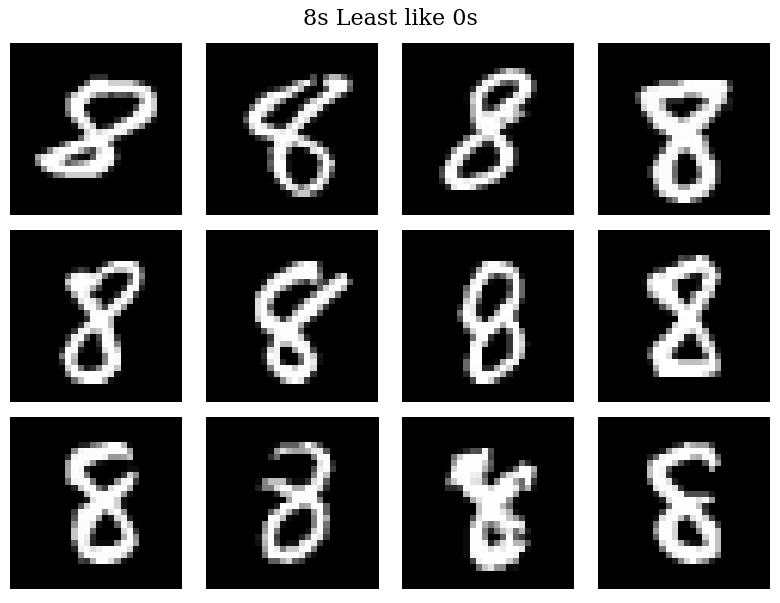}   
    \end{subfigure}
        \hspace{0.5cm}
        \centering
    \begin{subfigure}{0.4\textwidth}
        \centering
     \includegraphics[width=\textwidth]{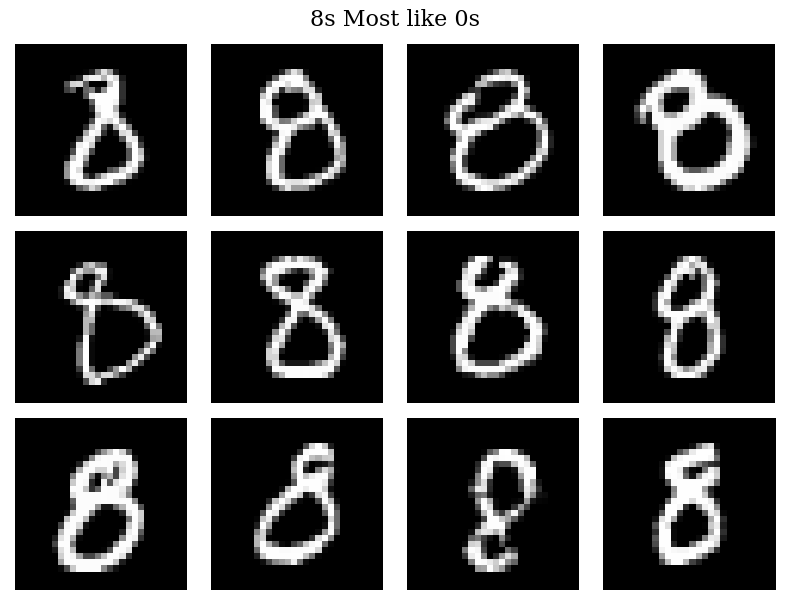}   
    \end{subfigure}
    \caption{MNIST images correctly classified (left) and misclassified (right) with a CNF trained on eights in contrast to zeros.}
    \vspace{-0.5cm}
    \label{fig:mnist_class_eights}
\end{figure}

\begin{figure}[t!]
    \centering
    \begin{subfigure}{0.4\textwidth}
        \centering
        \includegraphics[width=\textwidth]{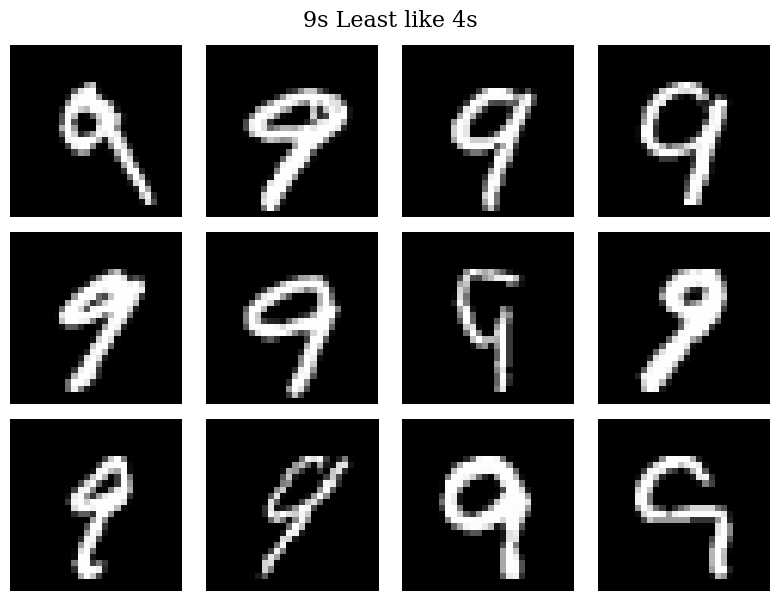}   
    \end{subfigure}
        \hspace{0.5cm}
        \centering
    \begin{subfigure}{0.4\textwidth}
        \centering
     \includegraphics[width=\textwidth]{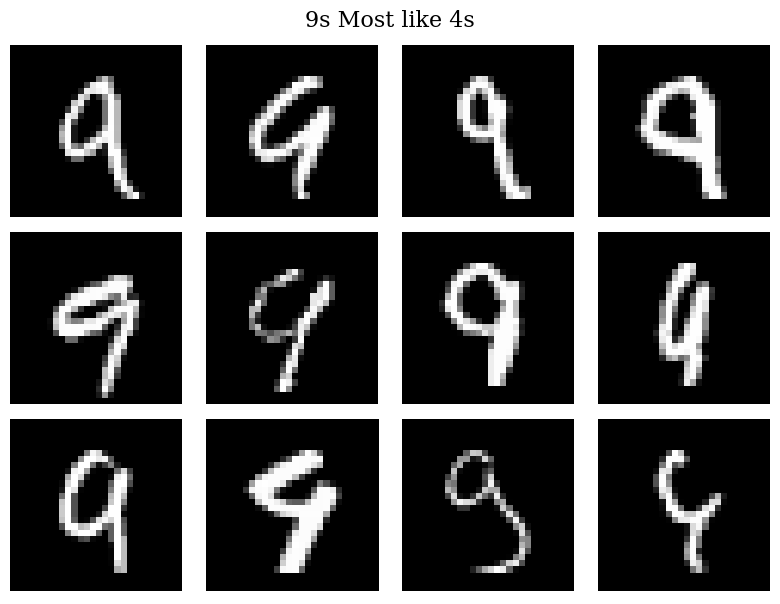}   
    \end{subfigure}
    \caption{MNIST images correctly classified (left) and misclassified (right) with a CNF trained on nines in contrast to fours.}
    \vspace{-0.5cm}
    \label{fig:mnist_class_nines}
\end{figure}

For a discriminative test, we train two CNFs, one with \(\mathcal D_s=\{\,\text{all eights}\,\}\) and \(\mathcal D_b=\{\,\text{all zeros}\,\},\) and one trained with \(\mathcal D_s=\{\,\text{all nines}\,\}\) and \(\mathcal D_b=\{\,\text{all fours}\,\}\), both with $c=2$. For each, we compute the discrimination function $\Phi = \frac{p_\theta^{(s)}}{1+p_\theta^{(s)}}$ on the test portion of the $\mathcal D_s$  dataset and let $\Phi >.5$ be our boundary for classification. We plot correctly classified events with the highest scores and incorrectly classified events with the lowest scores in Figure \ref{fig:mnist_class_eights} for the CNF trained with eights and zeros, and in Figure \ref{fig:mnist_class_nines} for the CNF trained with nines and fours. Since the CNF learns the entire distribution of one class of images, we can interpret classified/misclassified images as most like or least like the contrastive class. As seen in Figure \ref{fig:mnist_class_eights}, eights that are least like zeros (i.e have the highest values of $\Phi$) tend to have two equaly-sized lobes, and ones that are most like zeros have asymmetric lobes, with some visually approaching a zero as one of the lobes becomes very small. For Figure \ref{fig:mnist_class_nines}, the nines that are least like fours are generally closed on the top, while the nines that are most like fours are generally more open on the top, some closely resembling fours.  


\section{Details on the HiggsML Uncertainty Challenge}
\label{app:higgschal}
Here we summarize the relevant physics, the features that compose events in the dataset, the set of nuisance parameters, and the scoring for the FAIR Universe – HiggsML Uncertainty Challenge as described in Ref.~\cite{bhimji2024fairuniversehiggsmluncertainty}, to which we refer the reader for further details.
\subsection{Relevant physics background}
The challenge is based on high-energy proton–proton collisions at the Large Hadron Collider (LHC) and the subsequent classification of Higgs boson decays in the ATLAS detector.

\paragraph{Proton–proton collisions and Higgs production}
At the LHC, beams of protons are accelerated to near the speed of light and made to collide at a center-of-mass energy of 13 TeV.  Each collision (``event'') can produce a variety of particles, including the Higgs boson, through the conversion of kinetic energy into mass, as described by Einstein’s relation $E=mc^2$. Once produced, the Higgs bosons decay almost immediately through many decay channels. The challenge focuses on the decay channel $H \to \tau^+ \tau^-$ where one $\tau$ particle decays leptonically ($\tau\to \ell\nu\nu$) and the other hadronically ($\tau\to \text{hadron}+\nu$). Here, the lepton $\ell$ and the hadron are visible in the detector, but the neutrinos $\nu$ are invisible; as a result, the full kinematic features of the event are unavailable to the experiment. The events may also be accompanied by 0, 1, or 2 ``jets,'' sprays of collimated particles which are unrelated to the Higgs boson production but are produced as a result of secondary processes during the collision. The dataset for the challenge is composed of millions of simulated collision events, containing both signal $H \to \tau^+ \tau^-$ events and a variety of background events with the same visible final states.

\paragraph{The detector and the data features}
The ATLAS detector records the energy and direction of leptons, hadrons, and jets, which may be summarized by a momentum 3-vector $\vec{p}$ for each visible particle. Neutrinos escape detection, but their presence is inferred from the conservation of momentum in the direction transverse to the beams, via the computation of the missing transverse momentum vector
\begin{equation}
    \vec{P}_T^{\rm miss}
= \vec{0}- \sum_{i}\,\vec{P}_{T}^i \in \mathbb{R}^{2},
\end{equation}
where the sum runs over all reconstructed objects in the event. The kinematic information for each event (8 momentum coordinates for 0-jet events, 11 momentum coordinates for 1-jet events, and 14 momentum coordinates for 2-jet events), plus the number of jets and the sum of the transverse momenta for all jets in the event, comprise the primary features. Various nonlinear combinations of the primary features (described in detail in \cite{bhimji2024fairuniversehiggsmluncertainty}), corresponding to physically-motivated observables which have proved useful in collider physics analyses, bring the total to at most 28 features $F_i$ per event. These features are the only data accessible to participants; importantly, participants do \textit{not} have access to the true momentum vectors of the un-detected neutrinos.

\paragraph{Scientific motivation}
Precise measurement of the Higgs boson's properties -- such as the production rate (i.e., the signal strength $\mu$) and its couplings to other particles --tests the Standard Model and can be used to search for new physics.  The $\tau^+\tau^-$ channel is specifically important because it meausres the coupling strength of the Higgs to the heaviest lepton. The large simulated dataset and realistic detector effects the benchmark provides enable the development of methods that can robustly extract the rare signal in the presence of complex backgrounds, laying the groundwork for future precision studies at the LHC and beyond~\cite{bhimji2024fairuniversehiggsmluncertainty}. 

\subsection{Nuisance parameters}
The challenge introduces six nuisance parameters, \(\alpha_{\text{tes}}\), \(\alpha_{\text{jes}}\), \(\alpha_{\text{soft\_met}}\), \(\alpha_{\text{ttbar\_scale}}\), \(\alpha_{\text{diboson\_scale}}\) and \(\alpha_{\text{bkg\_scale}}\), which model systematic distortions in the data. In each pseudoexperiment, each of these parameters is drawn the distribution in Tab.~\ref{tab:nuisance-params}.

\begin{table}[ht]
  \centering
  \caption{Nuisance parameters: Gaussian (or log-normal) priors and clipping ranges.}
  \label{tab:nuisance-params}
  \begin{tabular}{lccc}
    \hline
    Parameter & Prior mean & Prior width \(\sigma\) & Clipping range \\
    \hline
    \(\alpha_{\text{tes}}\)           & 1.00 & 0.01 & [0.90, 1.10] \\
    \(\alpha_{\text{jes}}\)           & 1.00 & 0.01 & [0.90, 1.10] \\
    \(\alpha_{\text{soft\_met}}\)     & 0.00 & 1.00 & [0.00, 5.00] \\
    \(\alpha_{\text{ttbar\_scale}}\)  & 1.00 & 0.02 & [0.80, 1.20] \\
    \(\alpha_{\text{diboson\_scale}}\)& 1.00 & 0.25 & [0.00, 2.00] \\
    \(\alpha_{\text{bkg\_scale}}\)     & 1.00 & 0.001& [0.99, 1.01] \\
    \hline
  \end{tabular}
\end{table}

\paragraph{Effects on kinematic features}
Because the particle collisions take place at relativistic speeds, the fundamental objects which govern the signal and background distributions are relativistic 4-vectors $P = (E, \vec{P})$ where $E$ and $\vec{P}$ are the particle energy and momentum, respectively. A massless particle must satisfy $E^2 = \vec{P} \cdot \vec{P}$, but nuisance parameters in the form of mis-measurement can distort this relationship. The first three nuisance parameters $\alpha_{\text{tes}}$,  $\alpha_{\text{jes}}$, and $\alpha_{\text{soft\_met}}$ parameterize these distortions. For each event,
\begin{equation}
      P_{\text{had}}^{\rm biased} = \alpha_{\text{tes}}\; P_{\text{had}}, 
  \quad
  P_{\text{jet}}^{\rm biased} = \alpha_{\text{jes}}\;P_{\text{jet}}
\end{equation}
model mis-calibration of the hadronic and jet energy scales, respectively.  
These rescalings also shift the missing transverse momentum via
\begin{equation}
  \vec{P}_{T}^{\rm miss,\,biased}
  = \vec{P}_{T}^{\rm miss}
    +\bigl(1-\alpha_{\text{tes}}\bigr)\vec{P}_{\text{T, had}}
    +\bigl(1-\alpha_{\text{jes}}\bigr)\bigl(\vec{P}_{\text{T, jet,1}}+\vec{P}_{\text{T, jet,2}}\bigr),    
\end{equation}
and all changes propagate to the non-linear transformations that define the 28 features.  Finally, $\alpha_{\text{soft\_met}}$ smears the missing transverse momentum measurement by adding random Gaussian noise  $\mathcal{N}(0,\alpha_{\text{soft\_met}})$ to $\vec{P}_{T}^{\rm miss}$.

\paragraph{Impact on event weights}
The parameters \(\alpha_{\text{bkg\_scale}}\), \(\alpha_{\text{ttbar\_scale}}\) and \(\alpha_{\text{diboson\_scale}}\) modify the event weights (i.e. the frequency of some event) of the three sub-processes which comprise the background distribution:
\begin{equation}
    \begin{split}
         w' = \alpha_{\text{bkg\_scale}}\times w \quad
  (\text{for }Z\to\tau\tau \text{ background}),\\
  w' = \alpha_{\text{bkg\_scale}}\times\alpha_{\text{ttbar\_scale}}\times w 
  \quad(\text{for }t\bar{t} \text{ background}),\\\quad
  w' = \alpha_{\text{bkg\_scale}}\times\alpha_{\text{diboson\_scale}}\times w
  \quad(\text{for diboson background})
    \end{split}
\end{equation}
 altering the composition of overall background distribution without changing kinematic distributions.

\subsection{Metrics and scoring}
\label{app:metric}
Participants submit a method that, for each test set, returns a 68.27\% Confidence Interval (CI) for the Higgs signal strength \(\mu\).  The performance is evaluated over \(N_{\rm test}\) independent test sets, each generated with random values of \(\mu\) and the six systematic biases.

\paragraph{Precision and coverage}
Two criteria are used for computing the performance of a method. The first is precision, quantified by the average CI width
  \begin{equation}
    w \;=\; \frac{1}{N_{\rm test}} \sum_{i=1}^{N_{\rm test}} \bigl|\mu_{84,i} - \mu_{16,i}\bigr|\;,
    \label{eq:width}
  \end{equation}
  The second is coverage, the fraction of experiments in which \(\mu_{\rm true}\) lies within the CI:
  \begin{equation}
    c \;=\;\frac{1}{N_{\rm test}}\sum_{i=1}^{N_{\rm test}}
      [1 \text{ if }\bigl(\mu_{\rm true,i}\in[\mu_{16,i},\,\mu_{84,i}]\bigr) \text{ else 0}].
    \label{eq:coverage}
  \end{equation}
Deviations of \(c\) from the nominal 68.27\% are penalized via a piecewise function \(f(c)\), where \(\sigma_{68}=\sqrt{(1-0.68)\times0.68/N_{\rm test}}\) :
\begin{equation}
      f(c) = 
  \begin{cases}
    1, & 0.68 - 2\sigma_{68} \le c \le 0.68 + 2\sigma_{68},\\[6pt]
    1 + \biggl|\dfrac{c - (0.68 - 2\sigma_{68})}{\sigma_{68}}\biggr|^{4}, 
      & c < 0.68- 2\sigma_{68},\\[8pt]
    1 + \biggl|\dfrac{c - (0.68 + 2\sigma_{68})}{\sigma_{68}}\biggr|^{3}, 
      & c > 0.68 + 2\sigma_{68}.
  \end{cases}
\end{equation}
The cubic scaling for over-covering vs the quartic scaling for under-covering reflects the conservative preference for overestimating uncertainties as opposed to underestimating them.   
\paragraph{Final Score}
The overall \emph{coverage score} used to benchmark the method combines precision and coverage as
\begin{equation}
  \mathrm{score} \;=\; -\ln\bigl[(w + \varepsilon)\,f(c)\bigr]\;,
  \label{eq:final_score}
\end{equation}
where \(\varepsilon = 10^{-2}\) regularises against vanishing intervals. A higher score indicates both coverage at the correct rate while achieving as small an interval as possible. 

\section{Alternative likelihood-based method}
\label{app:np_method_overview}
 Here we describe a different method we developed that attempts to learn directly the likelihood in Eq. \ref{eq:nophilikli}. In this procedure, the event selection and pre-processing are \emph{identical} to the procedure described in Sec. \ref{sec:method_overview} as is the Neyman Construction once $\hat{\mu}$ is estimated. The main differences reside in the training and evaluation portion of our method.

\subsection{Training: DNN training with embedded nuisance parameters.}
Instead of training contrastive normalizing flows (CNFs), we pass the nuisance information \emph{directly} to the same DNN classifier:

\begin{enumerate}
\item Given a data point $\boldsymbol{x}$, let $\boldsymbol{\nu}^{true}\in\mathbb{R}^{k_\nu}$ and $\boldsymbol{\nu}^{fake}\in\mathbb{R}^{k_\nu}$ be two sets of different full 6D nuisance‑parameter vectors, where only $\boldsymbol{\nu}^{true}$ is the true set s.t $\boldsymbol{x}\sim p_{s/b}(\boldsymbol{x}|\boldsymbol{\nu}^{true})$.
\item During training, we append \emph{both} copies to each training event, and let the ordering depend on the label:
\begin{equation}
\tilde{\mathbf{x}}=\begin{cases}
\bigl(\mathbf{x},\boldsymbol{\nu}^{true},\boldsymbol{\nu}^{fake}\bigr) & \text{if the event is signal},\\
\bigl(\mathbf{x},\boldsymbol{\nu}^{fake},\boldsymbol{\nu}^{true}\bigr) & \text{if the event is background}
\end{cases}
\end{equation}
\item The same two‑headed DNN (one head for 1‑jet, one for 2‑jet events) with binary cross‑entropy loss is trained on this augmented input.
Thus, the network implicitly learns how variations of $\boldsymbol{\nu}$ shift the score distribution for each class.
\end{enumerate}

\subsection{Evaluation}
In this procedure, training is simpler, replaced with complexity in the evaluation procedure.
\paragraph{Nuisance‑parameter optimization}
At test time, we exploit the fact that all test data is overwhelmingly background (i.e., $f_s\ll1$). We freeze the \emph{signal} nuisance block in the input to a reference value $\boldsymbol{\nu}^{ref}$ (for example, the mean setting of $\boldsymbol{\nu}$) and vary the \emph{background} block $\boldsymbol{\nu}^{true}$ to \emph{minimize the sum of DNN scores} (i.e., to classify everything as background). That is, we first let:
\begin{equation}
    \tilde{\mathbf{x}} = \bigl(\mathbf{x},\boldsymbol{\nu}^{ref},\boldsymbol{\nu}^{test}\bigr)
\end{equation}
for all test points and we scan over $\boldsymbol{\nu}^{test}$ to find
\begin{equation}
\hat{\boldsymbol{\nu}}=\arg\min_{\boldsymbol{\nu}^{test}}\;\sum_{k\in\text{test}} r_k(\mathbf{x}_k;\boldsymbol{\nu}^{test}),
\end{equation}
where $r_k$ is the network output for event $k$.  This is an unbinned MLE under the (accurate) assumption that almost all events are background.

\paragraph{Template construction with perturbed data.}
Using $\hat{\boldsymbol{\nu}}$ we perturb a susbet of \emph{training} events to obtain two score templates, $S_{\text{sig}}(r\,|\hat{\boldsymbol{\nu}})$ and $S_{\text{bg}}(r\,|\hat{\boldsymbol{\nu}})$, that reflect the best‑fit nuisances.

\paragraph{Signal‑strength extraction.}
We create a histogram of scores $\{r_k\}$ by giving all inputs the same set $\hat{\boldsymbol{\nu}}$ for both the signal and background blocks, i.e.:
\begin{equation}
    \tilde{\mathbf{x}}_k = \bigl(\mathbf{x}_k,\hat{\boldsymbol{\nu}},\hat{\boldsymbol{\nu}}\bigr)
\end{equation}
and, given the signal and background templates from above for the best fit $\hat{\boldsymbol{\nu}}$ we can compute the MLE estimate for $\mu$.

This strategy replaces the CNF‑based likelihood construction with a purely classifier‑driven loop: nuisance parameters are absorbed as explicit inputs, profiled by maximizing the network’s global response, and finally used to perturb the score templates that enable a fit for the signal strength $\mu$ and $\hat{\boldsymbol{\nu}}$. A Neyman construction is once again used to generate confidence intervals.

\end{document}